\documentclass[twoside,twocolumn,9pt]{article}
\usepackage{extsizes}
\usepackage[super,sort&compress,comma]{natbib} 
\usepackage[left=1.5cm, right=1.5cm, top=1.785cm, bottom=2.0cm]{geometry}
\usepackage{balance}
\usepackage{mathptmx}
\usepackage{sectsty}
\usepackage{graphicx} 
\usepackage{lastpage}
\usepackage[format=plain,justification=justified,singlelinecheck=false,font={stretch=1.125,small,sf},labelfont=bf,labelsep=space]{caption}
\usepackage{float}
\usepackage{fancyhdr}
\usepackage{fnpos}
\usepackage[english]{babel}
\addto{\captionsenglish}{%
  
}
\usepackage{array}
\usepackage{droidsans}
\usepackage{charter}
\usepackage[T1]{fontenc}
\usepackage[usenames,dvipsnames]{xcolor}
\usepackage{setspace}
\usepackage{siunitx}
\DeclareSIUnit{\molar}{M}
\usepackage[compact]{titlesec}
\usepackage{hyperref}
\usepackage[version=4]{mhchem}

\usepackage{epstopdf}

\definecolor{cream}{RGB}{222,217,201}

\begin{document}

\pagestyle{fancy}
\thispagestyle{plain}
\fancypagestyle{plain}{
\renewcommand{\headrulewidth}{0pt}
}

\makeFNbottom
\makeatletter
\renewcommand\LARGE{\@setfontsize\LARGE{15pt}{17}}
\renewcommand\Large{\@setfontsize\Large{12pt}{14}}
\renewcommand\large{\@setfontsize\large{10pt}{12}}
\renewcommand\footnotesize{\@setfontsize\footnotesize{7pt}{10}}
\makeatother

\renewcommand{\thefootnote}{\fnsymbol{footnote}}
\renewcommand\footnoterule{\vspace*{1pt}%
\color{cream}\hrule width 3.5in height 0.4pt \color{black}\vspace*{5pt}} 
\setcounter{secnumdepth}{5}

\makeatletter 
\renewcommand\@biblabel[1]{#1}            
\renewcommand\@makefntext[1]%
{\noindent\makebox[0pt][r]{\@thefnmark\,}#1}
\makeatother 
\renewcommand{\figurename}{\small{Fig.}~}
\sectionfont{\sffamily\Large}
\subsectionfont{\normalsize}
\subsubsectionfont{\bf}
\setstretch{1.125} 
\setlength{\skip\footins}{0.8cm}
\setlength{\footnotesep}{0.25cm}
\setlength{\jot}{10pt}
\titlespacing*{\section}{0pt}{4pt}{4pt}
\titlespacing*{\subsection}{0pt}{15pt}{1pt}

\fancyfoot{}
\fancyfoot[LO,RE]{\vspace{-7.1pt}\includegraphics[height=9pt]{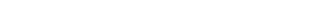}}
\fancyfoot[CO]{\vspace{-7.1pt}\hspace{13.2cm}\includegraphics{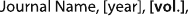}}
\fancyfoot[CE]{\vspace{-7.2pt}\hspace{-14.2cm}\includegraphics{head_foot/RF}}
\fancyfoot[RO]{\footnotesize{\sffamily{1--\pageref{LastPage} ~\textbar  \hspace{2pt}\thepage}}}
\fancyfoot[LE]{\footnotesize{\sffamily{\thepage~\textbar\hspace{3.45cm} 1--\pageref{LastPage}}}}
\fancyhead{}
\renewcommand{\headrulewidth}{0pt} 
\renewcommand{\footrulewidth}{0pt}
\setlength{\arrayrulewidth}{1pt}
\setlength{\columnsep}{6.5mm}
\setlength\bibsep{1pt}

\makeatletter 
\newlength{\figrulesep} 
\setlength{\figrulesep}{0.5\textfloatsep} 

\newcommand{\topfigrule}{\vspace*{-1pt}%
\noindent{\color{cream}\rule[-\figrulesep]{\columnwidth}{1.5pt}} }

\newcommand{\botfigrule}{\vspace*{-2pt}%
\noindent{\color{cream}\rule[\figrulesep]{\columnwidth}{1.5pt}} }

\newcommand{\dblfigrule}{\vspace*{-1pt}%
\noindent{\color{cream}\rule[-\figrulesep]{\textwidth}{1.5pt}} }

\makeatother

\twocolumn[
  \begin{@twocolumnfalse}
{
\includegraphics[width=18.5cm]{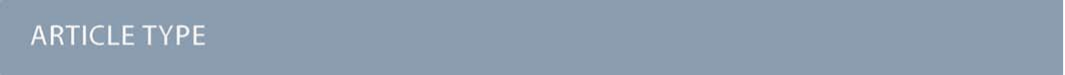}}\par
\vspace{1em}
\sffamily
\begin{tabular}{m{4.5cm} p{13.5cm} }

\includegraphics{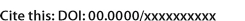} & \noindent\LARGE{\textbf{Surface Charge Relaxation Controls the Lifetime of Out-of-Equilibrium Colloidal Crystals}} \\
\vspace{0.3cm} & \vspace{0.3cm} \\

 & \noindent\large{Laura Jansen\textit{$^{a}$}, Thijs ter Rele$^{\ast}$\textit{$^{a}$}, and Marjolein Dijkstra\textit{$^{a}$}} \\

\includegraphics{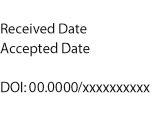} & \noindent\normalsize{Interactions between charged colloidal particles are profoundly influenced by charge regulation and charge renormalization, rendering the effective potential highly sensitive to local particle density. 
In this work, we investigate how a dynamically evolving, density-dependent Yukawa interaction affects the stability of out-of-equilibrium colloidal structures. 
Motivated by a series of experiments where unexpectedly long-lived colloidal crystals have suggested the presence of like-charged attractions, we systematically explore the role of charge regulation and charge renormalization. 
Using Poisson–Boltzmann cell theory, we compute the effective colloidal charge and screening length as a function of packing fraction. These results are subsequently incorporated into Brownian dynamics simulations that  dynamically resolve the evolving colloid charge as a function of time and local density. 
In the case of slow relaxation dynamics, our results show that incorporating these charging effects significantly prolongs the lifetimes of out-of-equilibrium colloidal crystals, providing an explanation for the experimental observation of long-lived crystals. 
These findings demonstrate that the interplay of surface charge dynamics and colloidal interactions can give rise to complex and rich nonequilibrium behavior in charged colloidal suspensions, opening new pathways for tuning colloidal stability through electrostatic feedback mechanisms.} \\

\end{tabular}

 \end{@twocolumnfalse} \vspace{0.6cm}

  ]

\renewcommand*\rmdefault{bch}\normalfont\upshape
\rmfamily
\section*{}
\vspace{-1cm}


\footnotetext{\textit{$^{a}$Soft Condensed Matter \& Biophysics, Debye Institute for Nanomaterials Science, Utrecht University, Princetonplein 1,3584 CC Utrecht, Netherlands; E-mail: 
t.r.terrele@uu.nl}}





\section{Introduction}\label{sec:introduction}
Suspensions of charge-stabilized colloids have attracted sustained interest in soft matter research for nearly a century, driven by their relevance in diverse applications ranging from industrial coatings to pharmaceutical formulations.\cite{matijevic-2008} The theoretical framework for understanding interactions in these systems is provided by the Derjaguin-Landau-Verwey-Overbeek (DLVO) theory, which describes the screened electrostatic colloidal interactions in an electrolyte through a distance-dependent Yukawa potential.\cite{verwey-1948, derjaguin-1993} Despite its conceptual simplicity, the DLVO potential has proven highly effective in describing the behavior of most charged colloidal systems and remains a central  tool in colloid science.\cite{Israelachvili2011, trefalt-2017} 

Since the 1980s, several experiments have cast doubts on the validity of DLVO theory for colloidal suspensions under low-salt conditions.\cite{yoshiyama-1984, yoshida-1990, ito-1988} These studies reported evidence of long-ranged like-charge attractions between colloids—an effect not predicted by the DLVO framework.  Observed phenomena suggesting such attractions include vapor–liquid phase separation,\cite{tata-1992} the formation of dilute voids,\cite{ito-1994,yoshida-1995,tata-1997} clustering of colloids,\cite{kubincova-2020, wang-2024} and colloidal crystals with unexpectedly long lifetimes.\cite{larsen-1996, larsen-1997} The origin of these apparent attractive interactions has yet to be resolved. 
It is important to note, however, that these findings remain controversial. Some results have proven difficult to reproduce,\cite{Palberg1994_tata} while others may stem from experimental artifacts, such as  out-of-equilibrium hydrodynamic effects caused by the proximity of the particles to a substrate.\cite{Squires2000_brenner}

Another intriguing phenomenon observed in charged-colloid suspensions is reentrant melting, where colloids undergo a phase transition from a crystal to a fluid phase upon increasing the density.\cite{royall-2006, kanai-2015} This behavior seems to originate from charge regulation mechanisms. Here, charge regulation is an umbrella term that includes a broad class of processes in which neither the surface charge nor the surface potential is constant.\cite{markovich-2016, ninham-1971} The specific form of charge regulation depends on the underlying charging mechanism, which can involve either ionizable surface groups\cite{ninham-1971} or the adsorption of ionic species.\cite{kanai-2015, royall-2006, everts-2016} Depending on the nature of the surface chemistry and solvent environment, the surface charge may increase\cite{hallett-2018} or decrease\cite{kanai-2015, royall-2006, everts-2016, zoetekouw-2006} with increasing colloid density. Furthermore, these charging mechanisms do not occur instantaneously, but are typically dynamic with finite timescales associated with charging processes.\cite{biswas-1998, avena-1999, krumina-2016, frens-1972}

{A related but distinct density-dependent phenomenon is charge renormalization, in which the bare colloid charge is replaced by a reduced effective charge that accounts for the counterions  condensed on the particle surface.\cite{alexander-1984, trizac-2003, belloni-1998} This effective charge enables the use of linearized Poisson-Boltzmann theories, while still capturing nonlinear screening effects. 
}
 As a result, both the colloid charge and the screening length become renormalized quantities.\cite{alexander-1984, trizac-2003} Charge renormalization is widely used to reconcile the significantly lower effective charges observed in experiments with the much higher bare charges typically obtained from titration measurements.\cite{gisler-1994, quesadaperez-1999, palberg-1995, von-grunberg-1999, rojas-ochoa-2008}

In this work, we employ Poisson–Boltzmann cell model calculations to determine the renormalized charge and screening length as a function of the packing fraction of the colloids,\cite{trizac-2003, zoetekouw-2006, everts-2016} following the approach of Zoetekouw \textit{et al.}\cite{zoetekouw-2006} These results serve as input for Brownian dynamics (BD) simulations  that explicitly account for the time- and density-dependent evolution of the colloid charge. Using these simulations, we investigate the structural lifetimes of  colloidal crystals in both cubic and planar slab geometries, focusing on how they are influenced by variations in charging dynamics. 

To study this theoretical model in a physically relevant setting, we chose parameters consistent with the experiments of Larsen and Grier.\cite{larsen-1996, larsen-1997} In their experiments,  polystyrene sulfate colloids with a diameter of $652 \pm 5 ~\textrm{nm}$ were used  at a low volume fraction of $\eta = 0.02$.\cite{larsen-1997} These particles carried a high surface charge due to strongly acidic sulfate surface groups, enabling them to be compressed into crystalline structures using an oscillating electric field. Upon removal of the field, the crystals reverted to a homogeneous fluid phase. Notably, superheated colloidal crystals prepared at relatively high ionic strength melted within approximately $10$ seconds, whereas those at lower ionic strength remained metastable for up to $30$ minutes, before reaching their equilibrium state.\cite{larsen-1996, larsen-1997} These long-lived structures also exhibited significant density variations, with differences as high as 70\%.\cite{larsen-1996}

We investigate these so-called superheated crystals from a new perspective, aiming to demonstrate not only our new   method  for incorporating  dynamic surface charge relaxation  but also  the impact  of density-dependent interactions on the stability of charged colloidal crystals in suspension. Using the  BD simulations described above, we systematically study how the  lifetime of a crystal depends on charge regulation and charge renormalization, with particular focus  on the rate at which these charging and decharging processes occur.  

We demonstrate that crystal lifetimes are  significantly enhanced by reducing the rate at which colloids equilibrate their surface charge, indicating that slow, density-dependent charging processes contribute to stabilizing the superheated crystal state. However, the simulated crystals do not exhibit the pronounced density differences as observed in experiments,\cite{larsen-1996} suggesting that additional mechanisms beyond our current model are needed to fully explain the experimental observations.

This paper is organized as follows. In Section \ref{sec:model_and_theory}, we introduce the experimental and theoretical background motivating  our Poisson-Boltzmann (PB) cell theory  calculations and discuss the physically relevant parameter regimes. Section \ref{sec:Methods} outlines the BD simulations and the machine-learning-based analysis techniques employed in this work. In Section \ref{sec:Results_total}, we present  the results of our PB cell theory calculations, which are then employed in BD simulations to investigate the structural lifetimes of superheated colloidal crystals in both  cubic and planar slab geometries, with the latter compared directly to experimental observations. 
Finally, we summarize our key findings and conclusions in Section \ref{sec:summary_and_conclusions}.

\section{Model and theory}\label{sec:model_and_theory}

\subsection{Interactions between Charged Colloids}
According to the classical DLVO theory, the effective interaction potential between two identical, homogeneously charged colloidal spheres of  radius $a$ and  total charge $Ze$ with $e$  the elementary charge, immersed in a solvent containing both coions and counterions with  dielectric constant $\epsilon$  and Debye screening length $\kappa^{-1}$,  is described by a purely repulsive screened-Coulomb (Yukawa) interaction \cite{verwey-1948,derjaguin-1993} 
\begin{equation}    \label{eq:hard_core_yukawa_potential}
    \beta U_{Yuk}(r) = Z^2 \lambda_B \left( \frac{\exp{(\kappa a)}}{1 + \kappa a} \right)^2 \frac{\exp{(-\kappa r)}}{r}, 
\end{equation}
where $r$ represents the center-of-mass distance between the  spheres, $\beta = 1/k_BT$ with $k_B$ Boltzmann's constant and $T$ the temperature, and $\lambda_B = e^2/4\pi\epsilon k_B T$ is the Bjerrum length. Note that we ignore the dispersion forces here.

In addition, we describe the hard-core repulsion between the colloids by a Weeks-Chandler-Andersen (WCA) potential \cite{weeks-1971}
\begin{eqnarray}
\label{eq;wca}
   \beta U_{WCA}(r) &=& \nonumber \\  & & \hspace{-15mm} \begin{cases}
4\beta\epsilon_{WCA}\left\lbrack \left(\frac{\sigma}{r} \right)^{12} - \left( \frac{\sigma}{r} \right)^6 + \frac{1}{4} \right\rbrack & r \leq 2^{\frac{1}{6}}\sigma,\\
0 & r > 2^{\frac{1}{6}}\sigma,
\end{cases}
\end{eqnarray}
where $\sigma=2a$ denotes the particle diameter with $a$ the particle radius, and $\epsilon_{WCA}$ is a parameter that sets the strength of the WCA potential. We use $\beta \epsilon_{WCA} = 40$, which has been used extensively in previous studies to model hard spheres.\cite{filion-2011} The total interaction potential between two colloids is simply the sum of the two potentials, $\beta U(r) = \beta U_{Yuk}+ \beta U_{WCA}(r)$. In this work, the range of the colloid interaction always exceeds $r>2^{1/6}\sigma$, so the specific form of the WCA potential serves primarily as a formal repulsive core rather than playing a significant physical role.

This study explores the effect of charge renormalization and charge regulation on the stability of out-of-equilibrium colloidal structures by taking into account a dynamically evolving Yukawa interaction that depends on the instantaneous local colloid density. Below, we describe how  charge renormalization and charge regulation are accounted for within the framework of a Poisson-Boltzmann cell theory. 

\subsection{Charge Renormalization and Charge Regulation}
\label{Poisson-Boltzmann}
{ Originally, Derjaguin, Landau, Verwey, and Overbeek derived a Yukawa-like interaction potential by linearizing the Poisson–Boltzmann equation and applying the linear superposition approximation, i.e., using single-sphere solutions of the linearized Poisson–Boltzmann (PB) equation to describe the interaction potential between two spheres. This provided a convenient analytical form for screened electrostatic interactions between charged colloids.\cite{verwey-1948, derjaguin-1993}
}
However, when the electric or surface potential of the colloids become sufficiently large, non-linear effects become significant and the linearization of the Poisson-Boltzmann equation and the linear superposition  break down. In such regimes, experimental studies have consistently shown that the effective colloid charge, obtained by fitting the interactions to a Yukawa potential, is significantly lower than the bare charge determined through titration, which directly probes the surface chemical groups.\cite{palberg-1995, quesadaperez-1999, gisler-1994} In other words, the bare charge overestimates the charge needed in the Yukawa potential to accurately describe phase behavior. To resolve this, Alexander \textit{et al.} introduced the concept of charge renormalization, in which the long-range asymptotic decay of the full non-linear PB solution is matched to a Yukawa form with an effective charge.\cite{alexander-1984, trizac-2003} This approach captures non-linear screening effects while retaining the analytical simplicity of the Yukawa potential, thereby improving the accuracy of theoretical predictions for colloidal interactions and phase behavior. 

In many theoretical treatments, the charge on a colloidal surface is assumed to be constant and fixed. However, this assumption often fails to capture the behavior observed in realistic experimental systems, where  the origin of surface charge must be explicitly considered. Typically, the charge  arises from the dissociation of chemical groups on the colloidal  surface in a polar solvent, such as water, resulting in a net surface charge. As a result, the colloid charge is not a fixed quantity but a dynamic one that can vary with system parameters such as temperature, pH, colloid packing fraction, and background ion concentration. This phenomenon, known as charge regulation, must be taken into account  to accurately describe colloidal interactions under realistic experimental conditions.

\subsubsection{Poisson-Boltzmann Cell Theory}

To account for non-linear screening and charge regulation effects, we employ the spherical cell approximation, which allows us to solve the full non-linear Poisson-Boltzmann (PB) equation. In this framework, a system of $N$ colloidal particles is modeled by partitioning the volume into identically sized spherical Wigner-Seitz cells, each containing a single colloid at its center.\cite{deserno-2001} The cell is characterized by a radius $R$, which is related to the particle size $a$ and the colloid packing fraction $\eta$ through the relation $\eta=\eta_{max}(a/R)^3$. Here, $\eta_{\mathrm{max}} \simeq 0.74$ represents the maximum packing fraction of a face-centered cubic (FCC) or hexagonal close-packed crystal of hard spheres, as used in Ref. \citenum{zoetekouw-2006}. While this convention of introducing a maximum packing fraction is not universally adopted in the literature,\cite{trizac-2003, everts-2016, smallenburg-2011} it can be reverted to the standard form by setting $\eta_{max}=1$. This reduces the complex many-body problem to a one-body problem with spherical symmetry, enabling the numerical solution of the non-linearized PB equation inside a single spherical cell. Within the Poisson-Boltzmann framework, the ion distributions follow a Boltzmann distribution $n_\pm(r) = n^{\infty} \exp{\left(\mp \phi(r) \right)}$, where $\phi(r) = e\psi(r)/k_B T$ is the dimensionless electric potential, $r$ is the radial distance from the center of the colloid, and $n^{\infty}$ is the bulk concentration of  coions or counterions.\cite{ohshima-2012} Substituting the total charge density $\rho_c (r) = e n_+ (r) - e n_- (r)$ into the Poisson equation from electrostatics $\nabla^2 \psi(r) = -\rho_c(r)/\epsilon$, leads to the non-linear PB equation 
\begin{equation}
    \label{eq:Non-linear_Poisson-Boltzmann}
 \nabla^2 \phi(r) = \kappa^2 \sinh{\phi(r)}, 
\end{equation}
where $\kappa=\sqrt{8\pi \lambda_B n^{\infty}}$ 
is the inverse Debye screening length.\cite{ohshima-2012, griffiths-2017}

To determine the renormalized charge, we follow the approach by Alexander {\em et al.} \cite{alexander-1984} and numerically solve the non-linear Poisson–Boltzmann (PB) equation (\ref{eq:Non-linear_Poisson-Boltzmann}) under appropriate boundary conditions. 
Since the PB equation is a second-order differential equation, two boundary conditions are required to obtain a unique solution. The first boundary condition, 
\begin{equation}
    \label{eq:cell_neutral}
   \left.  \frac{d\phi(r)}{dr}  \right|_{r=R} = 0,
\end{equation}
ensures total charge neutrality within the spherical cell.\cite{griffiths-2017} The second boundary condition is imposed at the colloidal surface and reflects  the surface chemistry of the colloids. This condition can vary depending on whether the surface charge or surface potential is held fixed, or if charge regulation is taken into account. In this work, we focus on dissociation reactions of acidic groups, in line with the polystyrene sulfate colloids we use as a model system.\cite{larsen-1996, larsen-1997}
{This dissociation equilibrium is described by
\begin{equation}
    \mathrm{AH} \ce{<=>} \mathrm{A^-} + \mathrm{H^+},
\end{equation}
where $\mathrm{A^-}$ represents a  surface charge group on the colloid and $\mathrm{H^+}$ the released counterion. This 
reaction leads to a relation between surface charge and surface potential \cite{ninham-1971, biesheuvel-2004A}
\begin{equation}
     \label{eq:alpha_Bies}
    \frac{Z}{M} = \frac{1}{1 + 10^{\mathrm{p}K_{a}-\mathrm{pH}}\exp{\left( -\phi(a) \right)}},
\end{equation}
where $M$ is the number of surface groups, $Z$ is the bare surface charge and $\mathrm{p}K_{a}$ and $\mathrm{pH} = -\text{log}([\text{H}^+])$ are the negative logarithmic acid dissociation constant and the negative logarithmic hydrogen ion concentration, respectively.} {The derivation of this relation,
presented in Appendix A, follows directly from the chemical equilibrium conditions at
the colloid surface.}
The bare charge is related to the electric potential via Gauss' law,\cite{griffiths-2017, smallenburg-2011}  
{ forming the second boundary condition to the non-linear PB equation.} For strongly acidic surface groups, the assumption of a constant charge is equally viable.
{ Moreover, ion-exchange resins are used to replace stray ions with a dominant species. Thus, for all intents and purposes, the counterion concentration can be taken to be equal to $[\text{H}^+]$.}

\subsubsection{Linearized Poisson-Boltzmann Cell Theory}

Using the solution to the non-linear Poisson-Boltzmann equation, we can calculate the renormalized charge $Z^*$ and the renormalized inverse screening length $\tilde{\kappa}$. However, when transitioning from the non-linear to the linearized Poisson-Boltzmann equation, one must choose a reference point about which  the equation is linearized. Conventionally, this is done at the cell boundary $R$,\cite{smallenburg-2011, everts-2016} but one could also linearize around the mean electric potential $\phi_m$, resulting in \cite{trizac-2003, brito-2023}
\begin{equation}
    \label{eq:effective_brito}
    Z^* = \frac{a\tanh{\phi_m}}{\lambda_B}\frac{a^2\tilde{\kappa}^2 \left[(R/a)^3 - 1 \right]}{3},
\end{equation}
and
\begin{equation}
    \label{eq:kappa_mean}
    \tilde{\kappa}^2 = \kappa^2 \cosh{\phi_m},
\end{equation}
where the mean electric potential $\phi_m$ is given by
\begin{equation}
    \label{eq:mean_potential}
    \phi_{m} = \frac{4\pi}{V_f}  \int_{a}^{R}  \phi(r)\,r^2 dr. 
\end{equation}
Here, $V_f$ refers to the free volume  within the spherical Wigner-Seitz cell, which reads \cite{trizac-2003, brito-2023}
\begin{equation}
    V_f = \frac{4\pi R^3}{3}\left(1 - \frac{\eta}{\eta_{\mathrm{max}}}\right).
\end{equation}
In this work, we employ the  cell model linearized around the mean electric potential (Eq.(\ref{eq:effective_brito})), as it enables  a one-to-one mapping between $Z^*$ and $\tilde{\kappa}a$ over a broad range of packing fractions $\eta$. { This mapping simplifies the numerical methods concerned with updating the equations of motion  for these two parameters, as will be explained in more detail in Section \ref{sec:methods_BD}.} 
The theoretical framework and results  for the linearized cell model using the cell boundary $R$ are provided in Appendix B.

\subsection{Relaxation Dynamics of Surface Charge and Screening Length}\label{sec:counterion_cond}

Using Eqs. (\ref{eq:effective_brito}) and (\ref{eq:kappa_mean}), the renormalized surface charge $Z^*$ and inverse screening length $\tilde{\kappa}$ are evaluated as functions of the local colloid packing fraction $\eta$. These quantities form the basis for the out-of-equilibrium Brownian dynamics simulations, which employ  the Yukawa potential defined in Eq. (\ref{eq:hard_core_yukawa_potential}). To determine the local colloid packing fraction $\eta$,  the simulation volume is partitioned into Wigner-Seitz cells using the Voronoi tesselation, for instance via \textsc{voro++}.\cite{rycroft-2009} As the system evolves, the local packing fraction $\eta$---and consequently  the surface charge $Z^*$ and reduced screening parameter $\tilde{\kappa}a$ --- changes dynamically. Instead of  updating these quantities instantaneously, we introduce a time delay that accounts for the finite relaxation time between changes in the  chemical environment and the  relaxation to a new (renormalized and regulated) charge. This rate of change is governed  by the surface chemistry underlying  charge renormalization and charge regulation.

\subsubsection{A Microscopic Model for Charge Renormalization}

So far, we have treated charge renormalization in a coarse-grained manner. To develop a more  microscopic description, we draw parallels to Manning’s counterion condensation theory for polyions.\cite{manning-1969}

Manning’s 1969 theory predicts that counterions condense onto an infinitely long, thin line charge when the spacing between charged groups is smaller than the Bjerrum length.\cite{manning-1969} A later refinement classified condensed counterions as free, territorially bound, or site-bound.\cite{manning-1979} Territorially bound ions can move freely along the polyion surface, whereas site-bound ions are  more restricted. However, experiments  and simulations show that the site binding is short-lived, typically less than a few  nanoseconds.\cite{hinderberger-2009, lo-2008, granot-1982} For this reason, we only consider territorially bound and free counterions. 

Manning condensation was formulated  for cylindrical polyions, and counterion behavior depends strongly on geometry.\cite{manning-2007, tang-2022, gillespie-2014, belloni-1998} This geometric dependence is well illustrated by toy models  such as those of  Tang and Rubinstein.\cite{tang-2022} Despite this dependence on geometry,  counterion condensation and charge renormalization are closely related.\cite{diehl-2004, sanghiran-2000, gillespie-2014, gisler-1994} For example, Diehl and Levin \cite{diehl-2004} used  molecular dynamics simulations with monovalent counterions around a spherical colloid and recovered the renormalized charge as predicted by  Alexander \textit{et al.}~\cite{alexander-1984} In their approach,  condensation was identified by comparing  the potential energies and kinetic energies of the counterions, and  the renormalized charge was obtained by substracting the number of condensed counterions from the bare charge.

\subsubsection{Adsorption Kinetics for Counterion Condensation}\label{sec:The_evolution_equation for_an_adsorption_type_charging_mechanism}

In light of these observations, we propose that charge renormalization can be interpreted as an adsorption process. This interpretation is consistent with its established role in models of dynamical charge regulation.\cite{shulepov-1995, biesheuvel-2002, dukhin-1990} Adsorption-based frameworks are also widely used in studies of linear flexible polyions,\cite{muthukumar-2004, ghosh-2024, shi-2021, lo-2008, radhakrishnan-2022} and, to a lesser extent for globular polyions.\cite{nikam-2020} 

A wide range of models exists for describing adsorption kinetics, varying in  complexity \cite{ho-2000}. In the absence of  experimental data specific to our  system that could justify a more elaborate model,  we adopt a simple and widely used pseudo-first-order kinetic scheme \cite{ho-2000, simonin-2016, rudzinski-2006}
\begin{equation}
    \frac{d N_{ads} (t)}{dt} = k\left(N_{ads, eq}(\eta) - N_{ads}(t)\right),
    \label{eq:differential_N}
\end{equation}
where $N_{ads, eq} (\eta)$ is the equilibrium number of adsorbed counterions on a single colloid and $N_{ads}(t)$ is the number of adsorbed counterions at time $t$. 
{For simplicity, we assume that the adsorption rate $k$ is constant, independent of counterion concentration, and equal to the desorption rate.}
Assuming all counterions are monovalent, we can identify  the number of adsorbed counterions with the adsorbed charge, $N_{ads, eq} (\eta) = Z_{ads, eq} (\eta)$ and $N_{ads} (t) = Z_{ads} (t)$. Following Diehl and Levin, we define  the renormalized charge as $Z^* = Z - Z_{ads}$,\cite{diehl-2004} where $Z$ is the bare colloid charge arising  from  surface dissociation reactions, and $-Z_{ads}$ represents the reduction due to counterion adsorption. In our system, the bare charge can be treated as constant, since the  surface groups are strongly acidic and thus largely insensitive to  local density variations, as shown in Fig. \ref{fig:cube_multi_figure_1}(a). Hence, we set $dZ/dt = 0$, or equivalently, $Z = Z_{eq}$. Substituting this relation into the kinetic equation yields a  differential equation for the time evolution of the renormalized charge
\begin{equation}
    \label{eq:differential_Z}
    \tau \frac{d Z^* (t)}{dt} = Z^*_{eq}(\eta) - Z^*(t),
\end{equation}
where $\tau = 1/k$ represents the characteristic  timescale over which  the colloidal surface charge relaxes  towards equilibrium with its local  chemical environment. Hence, the time evolution of $Z^*$ follows from first-order adsorption kinetics.  To maintain consistency, we impose that the time evolution of the inverse screening length $\tilde{\kappa}$, as defined in Eq. (\ref{eq:effective_brito}), mirrors that of the renormalized charge by making use of the  one-to-one mapping between $Z_{eq}^*$ and $\tilde{\kappa}a$, { as is discussed in Section \ref{sec:methods_BD}}.
{Here, we follow  a framework similar to that of Boon
{\em et al.},\cite{Wboon-2023} who observed that 
the equilibration time of the electric double layer is typically much shorter than both the colloidal diffusion time and the timescales of most surface-charge reaction processes. This slow-reaction assumption allows us to describe the double layer in quasi-equilibrium. Consequently, we can employ the equilibrium DLVO potential at each instant, with the non-equilibrium effective charge $Z^*(t)$ and inverse screening length  $\tilde{\kappa}(t)$ serving as time-dependent inputs.}
{ Furthermore, we note that the same differential equation Eq. \ref{eq:differential_Z} has previously been used to model charging dynamics in aqueous nanochannels}.\cite{kamsma-2023}

{Finally, we emphasize that our adsorption model assumes a uniform surface charge density on the colloids. 
This simplification contrasts with more detailed charge-regulating Poisson-Boltzmann calculations, which can yield heterogeneous and asymmetric charge distributions between colloids, and in some cases predict like-charge attractions.\cite{majee-2020, ruixuan-2023} }

\subsubsection{Relaxation Time \texorpdfstring{$\tau$}{tau} for Charge Renormalization}\label{sec:relaxation_time_renormalization}

Lastly, we provide a tentative estimate of the timescale $\tau$ for charge renormalization, acknowledging the inherent difficulties and therefore restricting ourselves to a lower bound. We focus on the relaxation of the innermost region of the electric double layer during a Brownian encounter---when  two colloids approach  closely enough for  their double layers to overlap and become perturbed.\cite{lyklema-1980} Previous studies have shown that the diffuse part of the double layer relaxes rapidly and can be considered in near-instantaneous equilibrium under such conditions.\cite{lyklema-1980, weaver-1985,dukhin-1990} By contrast, the relaxation of condensed counterions involves a range of processes with different timescales and a strong dependence of interaction geometry.\cite{weaver-1985, lyklema-1980, lyklema-1999, biesheuvel-2002, krozel-1994}  

In our simple adsorption model, the characteristic timescale $\tau$ in Eq (\ref{eq:differential_Z}) should be comparable to, or longer than, that of  lateral counterion diffusion, in contrast to  diffusion-limited models.\cite{krozel-1994, weaver-1985} As a lower-bound estimate, we take $\tau$  to be the characteristic diffusion time along the colloid surface, obtained by  dimensional analysis \cite{weaver-1985, lyklema-1980}
\begin{equation}
    \tau_{\textrm{\small ion}} = \frac{a^2}{D_{\textrm{\small ion},c}},
\end{equation}
where $D_{\textrm{\small ion},c}$ is the diffusion coefficient of  ions in the condensed layer and $a$ the colloid radius. It is worth emphasizing that the particular nature of the counterion association  with the  surface may result in  $\tau \gg \tau_{\textrm{\small ion}}$. 

Using the diffusion coefficient of $H^+$ in bulk water as an estimate for $D_{\textrm{\small ion},c}$, we obtain a lower bound of $\tau_{\textrm{\small ion}} \approx 0.01 ~\textrm{ms}$. Experimental data suggests  that the diffusion coefficient in the condensed layer may be comparable to its bulk value.\cite{lyklema-2002, lyklema-1999, lyklema-1998_langmuir} However, the structure of the electric double layer is complex, and  the diffusion coefficients in the innermost layer remain debated.\cite{hartkamp-2018} In the presence  of significant energy barriers attributed to lateral diffusion, the effective diffusion constant may be orders of magnitude smaller than in bulk.\cite{weaver-1985}

Given these uncertainties, we also explore relaxation times $\tau$ that are  significantly larger than $\tau_{\textrm{\small ion}} \approx 0.01 ~\textrm{ms}$. Additionally, systems with bulky counterions in nonpolar solvents, which are also actively studied,\cite{hussain-2013, gillespie-2014, hallett-2018} may exhibit even slower relaxation dynamics, justifying our consideration of a broad range of  $\tau$ values. 

\subsubsection{Generalization to Charge Regulation}

The time-evolution equation  for charge renormalization (Eq. (\ref{eq:differential_Z})) can also be interpreted as a simplified model for charge-regulating systems governed by ion adsorption,\cite{trefalt-2015} where the rate-limiting timescale is determined by the kinetics of charge regulation rather than renormalization. In this case, the slow response is attributed  to charge regulation dynamics, while the time dependence of charge renormalization itself is neglected. Charge regulation can be much slower  than particle diffusion, sometimes requiring timescales  on the order of minutes to hours to reach equilibrium.\cite{biswas-1998, krumina-2016, avena-1999, lis-2014, werkhoven-2018} 
To reflect this broader  applicability, we also consider relaxation times up to $\tau \approx 3 ~\textrm{s}$.

{Here, we assume that the relaxation time $\tau=1/k$ in the charging dynamics of  Eq. (\ref{eq:differential_Z}) is constant. In reality,  the relaxation time $\tau$ is expected to increase with higher counterion concentrations  $\left[\mathrm{H^+}\right]$.\cite{takae-2018} 
Nevertheless, the order of magnitude of $\tau$ is primarily determined by the specific system under consideration, as discussed in Section \ref{sec:counterion_cond}. For this study, we  adopt a constant-$\tau$ assumption, which provides a reasonable approximation for  the essential relaxation dynamics.} 
{We have added a discussion on the concentration dependence of $\tau$ and the rationale for employing the constant-$\tau$  model in Appendix C.}
{Still, incorporating the concentration dependence of $\tau$ represents a natural extension for future work. } 
{Another natural extension would entail the decoupling of the time evolution of $\tilde{\kappa}^*$ and $Z^*$, as these can progress through different timescales.}

\section{Methods}\label{sec:Methods}

\subsection{Brownian Dynamics with a Dynamic  Colloid Charge and Screening Length}\label{sec:methods_BD}

In this work, we investigate the melting behavior of charged colloids using  BD simulations. The time evolution of particle positions is evaluated using  the Langevin integration scheme, where we follow the integration method  proposed by Ermak and Yeh.\cite{ermak-1974} In this approach,  the position $\mathbf{r}_i$ of  particle $i$ at time $t + dt$ is given by
\begin{equation}
    \mathbf{r}_i (t + dt) = \mathbf{r}_i (t) + \sqrt{2 D dt} \mathbf{W}(t) - \frac{D}{k_B T} \nabla U dt, 
\end{equation}
where $dt$ is the integration time step, and  $\mathbf{W}(t)$ denotes a Wiener process that introduces stochastic noise corresponding to thermal fluctuations to the particles.
The diffusion constant $D$ determines the rate at which colloidal particles diffuse, setting the characteristic timescale $\tau_d = a^2/D$, known as the \textit{diffusion time} $\tau_d$. This timescale reflects how long it takes a particle to diffuse a distance comparable  to its own size $a$. Finally, the interparticle forces are derived from the interaction potential $U(r)$, which consists  of a Yukawa potential accounting for the screened electrostatic repulsions and a hard-core WCA potential describing the excluded volume, as defined in Eqs. (\ref{eq:hard_core_yukawa_potential}) and (\ref{eq;wca}), respectively. 
By employing this integration scheme, we assume particles are in the overdamped limit and  neglect hydrodynamic interactions. These approximations allow us to study macroscopic length and timescales. Nonetheless, hydrodynamic coupling under  confinement\cite{Squires2000_brenner} and in the presence of charge regulation\cite{takae-2018} is  highly non-trivial and represents an interesting direction for extending our model.

In the simulations, we used a time step of $dt = 8\cdot 10^{-6} \tau_d$. The interaction potential was truncated and shifted to zero at a cutoff distance $R_c = 8\sigma$, following the truncated and shifted force method,\cite{allen-2017} resulting in a negligible interaction strength of at most $\beta U(R_c) \approx 0.0001$.

The interaction potential depends on the renormalized charge $Z^*$ and the inverse screening length $\tilde{\kappa}$; we explained how to derive both of these parameters in Section \ref{sec:model_and_theory}. Here, the mean electrostatic potential $\phi_m$ for each packing fraction $\eta$ was calculated using the Simpson integration method, while the PB cell model was solved numerically using the \textsc{Scipy solve\_bvp} routine. Solutions  for $Z^*$ and $\tilde{\kappa}$ were obtained for the  packing fraction range $\eta \in [0,0.735]$, and  discretised into 4000 evenly spaced points. During the BD simulations, the values of $Z^*$ and $\tilde{\kappa}$ were determined as continuous  functions of $\eta$ using linear interpolation.

To determine the instantaneous renormalized charge $Z^*$ and inverse screening length $\tilde{\kappa}$ of each colloid during the simulation, we first compute its local packing fraction $\eta$  using the  Voronoi tesselation algorithm provided by \textsc{voro}++,\cite{rycroft-2009} which is computationally demanding.   
{Initially, the renormalized charge $Z^*$ and inverse screening length $\tilde{\kappa}$ of each  particle are set to their equilibrium values, $Z^*_{eq}$ and $\tilde{\kappa}_{eq}$, for the initial configuration.} 
To balance accuracy and efficiency, we update the equilibrium charge $Z^*_{eq}$ of each colloid via the Voronoi tesselation 
at a frequency of at least $1000$ times per relaxation time $\tau$. This  ensures  sufficient temporal resolution for systems with fast dynamics while reducing computational costs in more slowly relaxing systems. 
{ The charge $Z^*(t)$ and inverse screening length $\tilde{\kappa}(t)$ are updated at each simulation time step $dt$  by integrating Eq. (\ref{eq:differential_Z}). While a time-evolution equation for $\tilde{\kappa}$ could in principle be derived from Eq. (\ref{eq:effective_brito}),  this  would require knowledge of the time dependence of additional parameters in Eq. (\ref{eq:effective_brito}) and would depend on the point of linearization. Instead, we evolve only $Z^*(t)$ and ensure consistency by using a one-to-one mapping between $Z^*(t)$ and $\tilde{\kappa}(t)$ at each time step. This mapping is valid within a specific range of packing fractions and allows us to accelerate the simulations. 
In the regime where a one-to-one mapping between $Z^*$ and $\tilde{\kappa}a$ exists, the value of $\tilde{\kappa}a$ consistent with the time-evolved $Z^*$ is determined via  an effective packing fraction using  linear interpolation.} 
Simulations were carried out for a maximum duration of  $160\tau_d$. 

In Eq. (\ref{eq:hard_core_yukawa_potential}), a single renormalised inverse screening length $\tilde{\kappa}$ is used, even though  two interacting particles generally experience  different local densities  and thus   different   $\tilde{\kappa}$ values. To account for this asymmetry, we approximate the interaction between particles $i$ and $j$ using the average inverse screening length, defined as $ \tilde{\kappa}_{ij} = (\tilde{\kappa}_i + \tilde{\kappa}_j)/2$. This approximation is expected to be reasonable in regions with small density gradients, such as those considered in Section \ref{sec:results_plate}. More detailed approaches could incorporate the full ion configurations around each colloid to define an effective screening length, but such treatments are beyond the scope of the present work.   

\subsection{Identification of Crystalline Particles}\label{sec:fracFCC}
To determine when the colloidal  crystals have fully melted, 
we use a machine-learning (ML) scheme to identify which particles are in a crystalline structure. To this end, we first classify 
the local environment of each particle by calculating the Bond Order Parameters (BOPs).\cite{steinhardt-1983} These BOPs  are based on spherical harmonics $Y^m_l$ of order $l$, where $m$ is an integer between $m=-l$ and $m=l$.   For each particle $i$, we calculate 
\begin{equation}
    \label{eq:complex_quantity_to_qlm}
    q_{lm}(i) = \frac{1}{N_b(i)}\sum_{j=1}^{N_b(i)}Y^m_{l}({\bf r}_{ij}),
\end{equation}
where ${\bf r}_{ij}={\bf r}_j-{\bf r}_i$ is the distance vector between particle $i$ and $j$, and the summation runs over all $N_b(i)$ nearest neighbours of $i$. 
To distinguish more accurately between distinct thermodynamic phases, we calculate the averaged BOPS  defined by Lechner and Dellago,\cite{lechner-2008} where we average  $q_{lm}(i)$ over  the nearest neighbours and the particle itself \cite{lechner-2008}
\begin{equation}
    \label{eq:averaged_complex_quantity_to_qlm}
    Q_{lm}(i) = \frac{1}{N_b(i)+1}\left(q_{lm}(i) + \sum_{k=1}^{N_b(i)} q_{lm}(k) \right).
\end{equation}
By averaging over $-l\leq m \leq l$, we define a rotationally invariant BOP
\begin{equation}
    \label{eq:LD_BOPs}
    Q_{l}(i) = \sqrt{\frac{4\pi}{2l+1} \sum^l_{m=-l} |Q_{lm}(i)|^2}.
\end{equation}
We calculated these BOPs (Eq. (\ref{eq:LD_BOPs})) for $l\in[2, 11]$ using the \textsc{Python} package \textsc{Pyscal3},\cite{menon-2019} where the nearest neighbours were determined using a Voronoi tessellation.

The BOPs for all particles over the full simulation trajectory were  projected onto two principal components  using Principal Component Analysis (PCA). A Gaussian Mixture Model (GMM) with two components was then fitted to this reduced dataset, under the assumption that one component represents  the fluid phase and the other  the FCC crystal phase. From the GMM, we obtained  a crystallinity probability for each particle, which was summed across all particles and plotted as a function of time. To mitigate the influence of residual noise, we manually set the long-time crystallinity fraction  to zero, effectively removing a small percentage of false positives.

In Appendix E, we provide a more detailed description of this method and demonstrate its  robustness across various ML  training conditions. The aim of this procedure is not to accurately identify  the local structure of individual colloidal particles, but rather to capture  the overall structural evolution of the system. Notably, these systems contain a substantial number  of interface particles, which are notoriously difficult to classify.\cite{gispen-2025} 

\section{Results and discussion}\label{sec:Results_total}

\subsection{Effect of Charge Regulation and Charge Renormalization on Colloidal Interactions}\label{sec:results_charge_renormalization}

\begin{figure*}[ht]
\centering
\includegraphics[width=0.999\textwidth]{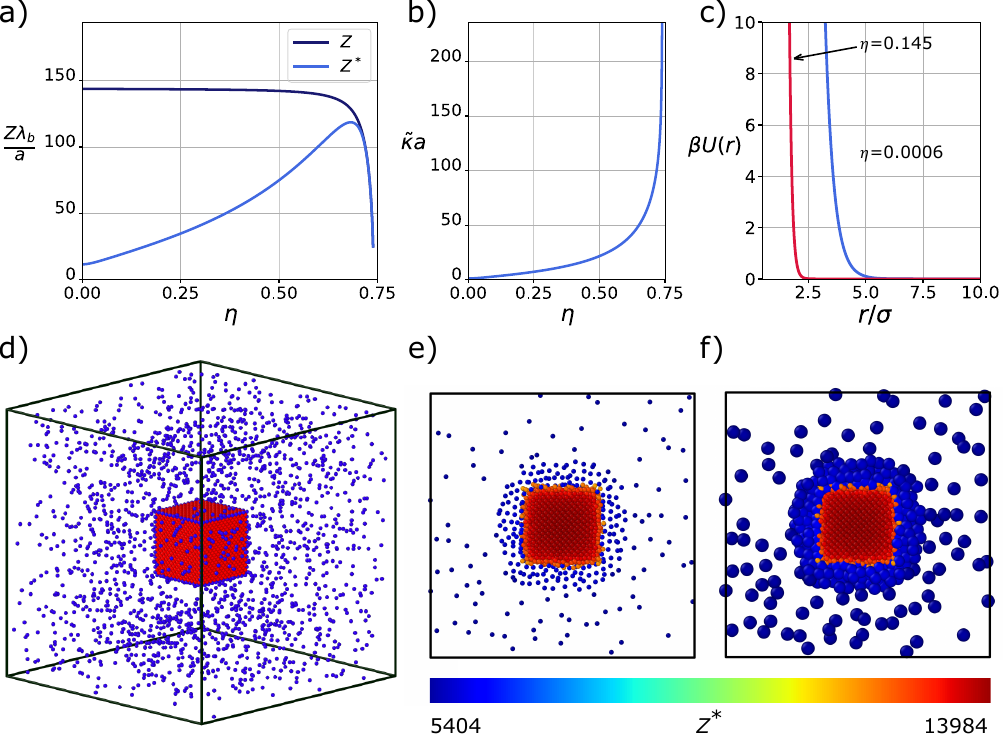}
\caption{\label{fig:cube_multi_figure_1}\textbf{Effect of Charge Regulation and Charge Renormalization on Colloidal Interactions. } 
 (a) Bare charge $Z$ (dark blue) and renormalized charge $Z^*$ (light blue) as  functions of packing fraction $\eta$.  (b) Renormalized  screening parameter $\tilde{\kappa} a$ as a function of $\eta$. (c) Total interaction potential $\beta U(r)$ between two colloids  as a function of center-of-mass distance $r$ for  two  packing fractions, $\eta=0.0006$ (blue) and 0.145 (red), representative of the initial gas  and crystal  phase, respectively. (d) Initial configuration used in the simulations consisting of a cubic crystallite surrounded by a dilute fluid phase. (e) Slice of thickness $5\sigma$ through  the center of the simulation box, showing the renormalized charge $Z^*$ of individual particles (see colorbar for color coding). (f) Same slice as in (e), but with  particle radii scaled by the local screening length according to $5/\kappa \sigma+0.7$. Panels (e) and (f) correspond to time  $t=2\tau_d$. The  parameters used in the calculations are adopted from Ref.~\citenum{larsen-1997}.}
\end{figure*}

Using the Poisson-Boltzmann cell theory as outlined in Section~\ref{Poisson-Boltzmann}, we calculate both the bare colloidal charge $Z$, which arises from charge regulation, and the renormalized charge $Z^*$, which accounts for charge renormalization, using Eq. (\ref{eq:alpha_Bies}) and (\ref{eq:effective_brito}), respectively. Additionally, we calculate the corresponding renormalized inverse screening length $\tilde{\kappa}$ using Eq.~(\ref{eq:kappa_mean}). These calculations are based on a simple dissociation boundary condition with  $M = 10^5$ the number of surface groups, $a = 465.5 \lambda_B$ the colloid radius, $\kappa a = 1.185$ the reduced screening parameter, and $[\text{H}^+]/K_{a} = 0.05$ the ratio between the bulk concentration of hydrogen and the acid dissociation constant. All parameters except $[\text{H}^+]/K_{a}$ are taken from the 1997 paper by Larsen and Grier,\cite{larsen-1997} where we use a Bjerrum length of $\lambda_B = 0.7 \si{\nano\meter}$. 
In Fig.~\ref{fig:cube_multi_figure_1}(a), we plot the bare colloid charge $Z$ (dark blue) and the renormalized charge $Z^*$ (light blue) as functions of the packing fraction $\eta$. We find that the bare charge $Z$ remains approximately constant over a broad range of packing fractions but decreases sharply as $\eta$ approaches the maximum packing fraction  $\eta_{max}$. In contrast, the renormalized charge  $Z^*$ exhibits a non-monotonic trend due to counterion condensation: it increases with $\eta$ at low packing fractions and begins to  decrease as $Z$ drops near $\eta_{max}$. The increase in $Z^*$ with $\eta$ has also been found in cell model calculations performed in the canonical ensemble.~\cite{hallett-2018}

Figure \ref{fig:cube_multi_figure_1}(b) shows the corresponding renormalized inverse screening length $\tilde{\kappa}$. We clearly observe that $\tilde{\kappa}$ increases rapidly with colloid packing fraction $\eta$ in this low-salt system. 
This increase in $\tilde{\kappa}$ results from the higher salt concentration at higher colloid density, as is described by the Gibbs-Donnan effect.\cite{donnan-1911}
Our choice for the ratio between the bulk concentration of hydrogen and the acid dissociation constant $[\text{H}^+]/K_{a}$ was limited by the numerical stability of our calculations. However, in Appendix B we show that the solution to the PB equation in this parameter space regime is largely insensitive to the exact value of the bare surface charge. Therefore, the chosen value of   $[\text{H}^+]/K_{a}$ is still expected to yield results that are representative of our model system.

The renormalized charge $Z^*$ and the inverse screening length $\tilde{\kappa}$ can be employed  in Eq. (\ref{eq:hard_core_yukawa_potential}) to define a dynamically evolving, density-dependent interaction potential. In Figure~\ref{fig:cube_multi_figure_1}(c), we illustrate this  interaction at two representative  packing fractions,   $\eta=0.0006$ and $\eta = 0.145$. This plot clearly shows that the Yukawa interaction is significantly more repulsive at low packing fraction. 
This reduction in  repulsion  at higher $\eta$ is primarily due to enhanced  electrostatic screening, reflected in the increase of  $\tilde{\kappa}$. Although the renormalized charge $Z^*$ also rises with density, its effect is outweighed by the stronger screening.

\subsection{Melting Behavior of a Cubic Crystallite}\label{sec:results_cube}

\begin{figure*}[ht]
\centering
\includegraphics[width=0.999\textwidth]{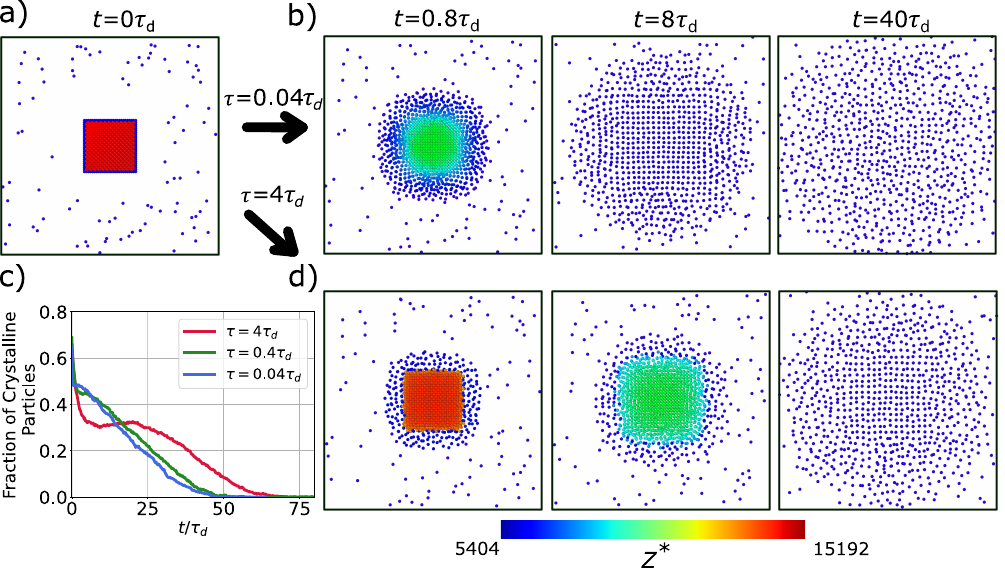}
\caption{\textbf{Melting Behavior of a Cubic Crystal with an FCC structure.} For clarity, only slices of thickness $6\sigma$ through the center of the simulation box are shown. The colors of the particles represent their renormalized charge $Z^*$, as indicated by the color bar.
(a) Slice of the initial configuration used in the  melting simulations. (b) Typical configurations from the melting simulation with relaxation time $\tau = 0.04 \tau_d$. (c) Time evolution of the fraction of particles identified  as  crystalline by our machine learning approach. (d) Typical configurations from the melting simulation with $\tau = 4 \tau_d$.}
\label{fig:snaps_tau2}
\end{figure*}

Using the dynamically evolving, density-dependent Yukawa potential described above, we study the lifetime of a small cubic crystallite with an FCC structure, surrounded by a dilute fluid phase,  for various values of the relaxation time $\tau$. 
The initial configuration, shown in Fig.~\ref{fig:cube_multi_figure_1}(d), consists  of a cubic crystallite of $8788$ particles arranged in an FCC structure with an initial packing fraction $\eta_i$, centered at the origin of the simulation box. This crystal is surrounded by $2000$ randomly placed particles in a fluid phase. To achieve the desired global packing fraction $\eta_g$, we adjust the total volume of the simulation box while keeping the initial packing fraction $\eta_i$ of the cubic crystal fixed. In all our simulations we use periodic boundary conditions in all three spatial directions. The parameters for all  configurations used in this work, including $\eta_g, \eta_i$, and the relaxation time $\tau$, are listed in Table~\ref{tab:cubes}, each identified by a code $B_i$ with {$i\in[1,16]$}. We first consider simulations $B_{9}$ - $B_{12}$, in which the crystal has an initial packing fraction of $\eta_i \simeq 0.231225$, the global packing fraction is  $\eta_g \simeq 0.004189$, and the fluid phase initially has a packing fraction of $\eta \simeq 0.0015$. 
{Note that, at equilibrium, the colloids form a homogeneous fluid phase at a global packing fraction of $\eta_g \simeq 0.004189$.   }

\begin{table}[H]
\centering
\setlength{\tabcolsep}{10pt}
\caption{Parameters for the initial configurations of the BD simulations on a cubic crystallite with an FCC structure surrounded by a dilute fluid phase.   The configurations are characterised by  a code $B_i$, a global packing fraction  $\eta_g$, the initial packing fraction of the crystal $\eta_i$, and the relaxation time $\tau$ used in the simulations.}
\label{tab:cubes}
\begin{tabular}{lllll}
\hline
Code   & $\eta_g $    & $\eta_i$ & $\tau/\tau_d$ \\ \hline
$B_1$ &  0.001047    &          0.057806              &        4      \\
$B_2$ &  0.001047    &          0.057806              &        0.4      \\
$B_3$ &  0.001047    &          0.057806              &        0.04      \\
$B_4$ &  0.001047    &          0.057806              &        0.004      \\
$B_5$ &  0.002094    &          0.115613              &        4      \\
$B_6$ &  0.002094    &          0.115613              &        0.4      \\
$B_7$ &  0.002094    &          0.115613              &        0.04      \\
$B_8$ &  0.002094    &          0.115613              &        0.004      \\
$B_{9}$ &  0.004189    &       0.231225                 &        4      \\
$B_{10}$ &  0.004189    &       0.231225                 &        0.4      \\
$B_{11}$ &  0.004189    &       0.231225                 &        0.04      \\
$B_{12}$ &  0.004189    &       0.231225                 &        0.004      \\
$B_{13}$ &  0.005236    &      0.404644                  &        4      \\
$B_{14}$ &  0.005236    &      0.404644                  &        0.4      \\
$B_{15}$ &  0.005236    &      0.404644                  &        0.04      \\
$B_{16}$ &  0.005236    &      0.404644                  &        0.004      \\
\hline
\end{tabular}
\end{table}

In the dense crystalline region, the effective charge of the particles is higher than in the surrounding dilute fluid, as shown in Fig. \ref{fig:cube_multi_figure_1}(e). However, the particle interactions are less repulsive in the dense region due to the increased screening, reflected by a larger value of  $\tilde{\kappa}a$. This effect is illustrated in Fig. \ref{fig:cube_multi_figure_1}(f), where the colloid radius gives an impression of the interaction range. Interestingly, particles in the gas phase appear effectively  larger to one another than those in the crystal, owing to the longer-range interactions at lower densities.

Typical configurations of the  system at different stages of its melting process are shown in Fig. \ref{fig:snaps_tau2}, illustrating how  the expansion of the crystal depends on the relaxation time $\tau$. At low values of $\tau$, like $\tau=0.04\tau_d$ with $\tau_d$ the diffusion time of the colloids (Fig. \ref{fig:snaps_tau2}(b)), the crystal rapidly expands into a dilute crystalline structure before eventually melting. In this regime,  the charging mechanism responds  almost instantaneously, so its time dependence is minimal. At longer relaxation times, $\tau = 4\tau_d$  (Fig. \ref{fig:snaps_tau2}(d)), the expansion proceeds much slower. After a time of $t = 8 \tau_d$, a dense FCC crystal still coexists with a surrounding dilute fluid phase. In contrast, for a short relaxation time of  $\tau = 0.04 \tau_d$,  a large portion of the simulation box is already occupied  by a dilute crystalline structure at the same time point. 

Two main factors contribute to the stability of the crystal. The first arises  from the relaxation dynamics within the dense region. Due to diffusion, particles tend to  move on average from dense to dilute regions. Since the range of repulsion  increases as the local  density decreases, this can trigger  a feedback loop that drives rapid  crystal expansion. However, when the relaxation time $\tau$ is large, this feedback loop is suppressed because the particles have not yet adjusted  to their new chemical environment. In other words, for small values of $\tau$, the particles closely follow the light blue curves in Fig. \ref{fig:cube_multi_figure_1}(a) and Fig. \ref{fig:cube_multi_figure_1}(b). In contrast, for  large  $\tau$,  particles can effectively take a variety of paths through parameter space---spanning $Z^*\lambda_b/a$, $\tilde{\kappa}a$, and $\eta$---rather than remaining on an instantaneous response curve. { This point is illustrated in Appendix D, where we study the time evolution of the charge and screening length of particles initially at the exterior and in the interior of the superheated crystal.}

\begin{figure*}
\includegraphics[width=0.99\textwidth]{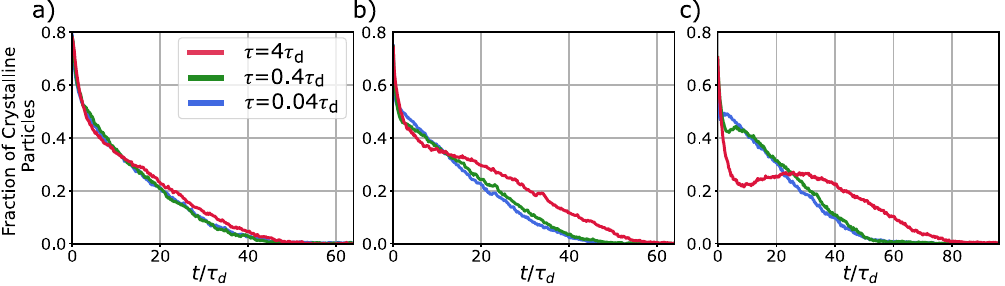}
\caption{\textbf{Melting Behavior of Cubic Crystallites.} Fraction of crystalline particles as a function of time $t/\tau_d$ for varying relaxation times  $\tau$ and different  initial packing fractions $\eta_i$ of the crystal and global packing fraction $\eta_g$: (a) $(\eta_i,\eta_g) = (0.057806,0.001047)$ from simulations $B_1, B_2, B_3$, (b) $(\eta_i,\eta_g) = (0.115613,0.002094)$ from simulations $B_5, B_6, B_7$, and (c) $(\eta_i,\eta_g) = (0.404644,0.004189)$ from simulations $B_{13}, B_{14}, B_{15}$. { In this dilute regime, the small variations in $\eta_g$ between  simulations have a negligible impact on the crystal lifetime.}} 
\label{fig:sup:density_com}
\end{figure*}

The second factor contributing to the stability of the crystal is the formation of an intermediate layer between the dense crystallite and the surrounding dilute fluid phase. When the charging dynamics is fast, i.e. small $\tau$,  particles at  the surface of the crystallite  quickly become more repulsive due to  their lower local density. As a result, they experience  strong outward repulsive forces that rapidly drive them away from the crystal center. In contrast, when the charging dynamics is slower, i.e. larger $\tau$, surface particles do not immediately adjust  to the reduced local density. Instead, they diffuse more gradually away from the crystallite, forming a dense, fluid-like layer. Although these particles have a lower local density than those inside the crystallite, their increased repulsion still plays a stabilizing role. This intermediate fluid layer acts as a barrier, exerting a counteracting force that resists further expansion of the crystal and thus enhances the stability and thereby the lifetime of the crystal. 
{ The strong repulsions in the intermediate fluid layer delay the breakup of the crystal, which ultimately melts due to weaker repulsions between colloids inside the crystal.}
{ Note that this effect strongly depends on the long screening length $\kappa^{-1}$ (resulting from the low ionic strength in the experiment),  since the repulsions in the dilute regions need to be long-ranged in order to stabilise the crystal.}
The effects of   increased stability of the crystal are shown  in Fig. \ref{fig:snaps_tau2}. At $\tau = 0.04 \tau_d$, the crystal has completely dissolved by  $t = 40 \tau_d$. In contrast, at   $\tau = 4\tau_d$,   crystalline regions remain visible at  the same time point. The stabilizing fluid layer surrounding the crystal is also clearly visible in this figure.

To enable a more precise comparison between  different values of $\tau$, we quantify the extent of crystal melting by measuring the fraction of  particles in an FCC-crystalline local environment. Fig. \ref{fig:snaps_tau2}(c) shows the time evolution of the  fraction of crystalline particles.   The method used to identify the crystalline particles is described  in Section \ref{sec:fracFCC} and detailed further  in  Appendix E. For each simulation, we calculated the BOPs for each particle.   The datasets from simulations $B_{9}$ - $B_{11}$ were projected onto the resulting principal components obtained by performing PCA on data from simulation $B_{12}$. Subsequently, a GMM was  trained on the projected datasets.  From Fig. \ref{fig:snaps_tau2}(c),  a clear distinction emerges between systems with a relaxation time  $\tau > \tau_d$ and those  with $\tau < \tau_d$. For $\tau < \tau_d$, the crystalline fraction gradually decreases until there is no crystal left around $t = 45 \tau_d$. In contrast, for $\tau > \tau_d$, the fraction of crystalline particles initially   plateaus, followed by  a gradual decrease, with complete melting occuring at approximately  $t = 65 \tau_d$. 

In addition, we examine how the initial packing fraction $\eta_i$ of the crystal influences its lifetime. We analyze three sets of simulations with varying initial packing fractions: $B_1$ - $B_4$ with $\eta_i=0.057806$, $B_5$ - $B_8$ with $\eta_i = 0.115613$, and $B_{13}$ - $B_{16}$ with $\eta_i = 0.404644$. For each set, PCA was performed using data from the simulation with  $\tau = 0.004\tau_d$. The remaining three  datasets in each set were projected onto the resulting principal components. A GMM was then trained and tested on each of the nine projected datasets. The resulting fraction of crystalline particles are presented in Fig. \ref{fig:sup:density_com}. Based on Fig. \ref{fig:sup:density_com}, we observe that  increasing the initial packing fraction $\eta_i$ enhances the lifetime of the crystal, 
{at fixed} relaxation time $\tau$. 
More specifically, the lifetime of the crystal increases progressively from Fig. \ref{fig:sup:density_com}(a) to Fig. \ref{fig:sup:density_com}(c). This effect is particularly pronounced at larger relaxation times $\tau$  (see the red curves). As the initial packing fraction $\eta_i$  increases,  the difference between the two regimes $\tau < \tau_d$ and $\tau > \tau_d$ becomes increasingly evident. At low initial packing fraction $\eta_i$ (Fig. \ref{fig:sup:density_com}(a)), the influence of $\tau$ on the melting rate is minimal. This is because  density-dependent interaction effects are weak in dilute systems, and introducing a time lag in the response of the interactions has little impact. In contrast, at higher initial packing fractions $\eta_i$  (Fig. \ref{fig:sup:density_com}(b) and Fig. \ref{fig:sup:density_com}(c)), the delayed response of the interactions plays a more important role. Here, larger values of  $\tau$ noticeably extend the lifetime of the crystal. This suggests that as the system is driven further out of  equilibrium, the crystal becomes more long-lived, and the delayed relaxation of the colloid surface chemistry plays an increasingly important role in the stabilization of the crystal. 

To convert the dimensionless timescales discussed in this section into physical units, we refer to the particle diffusion coefficients reported  by Larsen and Grier; either  $D = 0.12 ~\mathrm{\mu m^2 s^{-1}}$  \cite{larsen-1996} or $D = 0.15 ~\mathrm{\mu m^2 s^{-1}}$.\cite{larsen-1997} These values correspond to diffusion times of   either $\tau_d = 0.89 \mathrm{s}$ or $\tau_d = 0.71 \mathrm{s}$, respectively. Using the shorter diffusion time, $\tau_d = 0.71 \mathrm{s}$, the two relaxation times translate to $\tau = 0.04 \tau_d = 28 \mathrm{ms}$, and $\tau = 4\tau_d = 2.8 \mathrm{s}$, and the lifetimes of the crystal range from $32 \mathrm{s} -64 \mathrm{s}$. 

In summary, for this specific set of initial conditions, the charge relaxation time---approximately ranging from milliseconds to seconds---strongly influences the lifetime of a superheated crystal, since a larger relaxation time can double the crystal lifetime.
These results show that incorporating non-instantaneous charge regulation and charge renormalization effects into simulations of charged colloids enhances the stability and lifetime of crystalline structures.

\begin{figure*}
\centering
\includegraphics[width=0.999\textwidth]{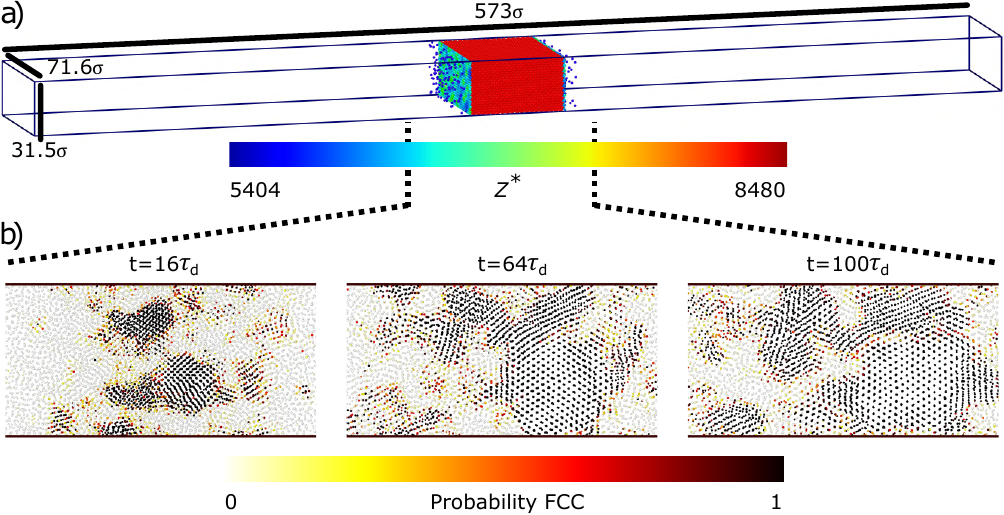}
\caption{\textbf{Melting Behavior of a Crystal with an FCC structure in a Slab Geometry.} (a) Initial configuration with particles colored according to their renormalized charge $Z^*$, as shown by the color bar. The crystal has an initial packing fraction of $\eta_i=0.089215$, and a global packing fraction of $\eta_g = 0.009$. The relaxation time is set to $\tau = 0.04 \tau_d$. This corresponds to simulation $P_3$ 
of Table \ref{tab:plates}. (b) Time evolution of the system during expansion of the crystal. The particle colors represent the  probability of  belonging  to the crystal phase. The images show  slices of thickness $10\sigma$, centered along the height  of the simulation box with dimensions of approximately $72\sigma \times 147\sigma$. For visual clarity, all particle radii are set to $0.7\sigma$.}
\label{fig:plates_main}
\end{figure*}

\subsection{Melting Behavior of a Crystal in a Slab Geometry}\label{sec:results_plate}

Next, we investigate a system with a geometry more closely resembling the experimental setup used by Larsen and Grier.\cite{larsen-1996, larsen-1997} While  the previous simulations of small compressed cubic crystals surrounded by a fluid phase allowed for  a quantitative analysis of the effect of the relaxation time $\tau$, the more complex geometry considered in the next simulations reveals that the specific details of the initial configuration also play a significant role in extending  the lifetime of superheated crystals.


The initial configuration of the particles is shown in Fig. \ref{fig:plates_main}(a). An FCC crystal consisting of 22000 particles with an initial packing fraction of $\eta_i$ is  positioned at  the centre of an elongated simulation box. We apply periodic boundary conditions in all three spatial directions, resulting in two planar crystal-gas interfaces. The surrounding dilute gas phase contains 194 randomly placed  particles. The total volume of the simulation box is adjusted to achieve a target global packing fraction $\eta_g$, while keeping $\eta_i$ of the crystal fixed. The height of the box is maintained at approximately  $31.5\sigma$,  corresponding to about $20.5~\mu\textrm{m}$ or $0.02~\textrm{mm}$---matching the height of the experimental setup used by  Larsen and Grier.\cite{larsen-1996} All relevant  parameters  are provided in  Table \ref{tab:plates}. 
A key difference between the simulations and experiments is, however, the use of periodic boundary conditions in the simulations. Furthermore, the simulations use a fully filled slab geometry, in contrast to the experiments, which consist of an unspecified number of layers. 
As a result, a direct quantitative comparison between the simulations and experiments is not feasible.  Nevertheless, in the following,  we highlight qualitative features that resemble those observed in the experimental system.

\begin{table}
\centering
\setlength{\tabcolsep}{10pt}
\caption{Parameters for the initial configurations of the BD simulations on a crystal in a slab geometry adjacent to a low-density gas phase, with the two phases separated by two planar interfaces. The configurations are characterised by a code $P_i$, a global packing fraction $\eta_g$,  initial packing fraction of the crystal $\eta_i$, and the relaxation time $\tau$ used in the simulations.}
\label{tab:plates}
\begin{tabular}{lllll}
\hline
Code   & $\eta_g $  & $\eta_i$ & $\tau/\tau_d$ \\ \hline
$P_1$ &  0.009 &         0.089215           &        4      \\
$P_2$ &  0.009 &        0.089215     &        0.4      \\
$P_3$ &  0.009 &         0.089215        &        0.04      \\
$P_4$ &  0.009 &        0.089215       &        0.004      \\
$P_5$ &  0.02 &       0.089213         &        4      \\
$P_6$ &  0.02 &          0.089213     &        0.4      \\
$P_7$ &  0.02 &         0.089213     &      0.04      \\
$P_8$ &  0.02 &          0.089213       &     0.004      \\
$P_9$ &  0.03 &         0.089213        &        4      \\
$P_{10}$ &  0.03 &      0.089213        &        0.4      \\
$P_{11}$ &  0.03 &     0.089213         &        0.04      \\
$P_{12}$ &  0.03 &      0.089213       &        0.004      \\
\hline
\end{tabular}
\end{table}

We first  examine a  system with an initial packing fraction of the crystal of  $\eta_i=0.089215$, and a global packing fraction of   $\eta_g = 0.009$. The relaxation time is set to $\tau = 0.04 \tau_d$. However, given the low  initial density of the superheated crystal, this value of $\tau$ is not expected to significantly influence  the out-of-equilibrium dynamics (see Fig. \ref{fig:sup:density_com}). 
The simulation presented here corresponds to simulation $P_3$ 
in Table  \ref{tab:plates}. 
To track the evolution of the crystal during expansion, we employ our Machine Learning  method for identifying  crystalline particles described in Section \ref{sec:fracFCC} and further detailed in  Appendix  E.  To obtain the FCC class probabilities shown in Fig. \ref{fig:plates_main}, we projected the BOPs from $P_3$ 
onto the first two principal components obtained from simulation $P_2$. 
A GMM was then trained and tested on these reduced BOPs. Fig. \ref{fig:plates_main}(b) illustrates the expansion of the crystal, with particles color-coded according to their probability of belonging to the  crystal phase. The color-coding was also validated against visual inspection of simulation trajectories in Supplementary Movies of the system.\cite{sm} Due to the large system size, the simulation could not be run until it reaches full equilibrium. 

We make the following observations when we run such a simulation. After an initial relaxation, the FCC crystal at the center of the box falls apart into irregularly shaped domains, separated by disordered particles, as shown  in the left frame of Fig. \ref{fig:plates_main}(b). The subsequent frames illustrate the time evolution of these ordered regions. Notably, islands of FCC-structured particles persist and even grow in some cases in the central region  of the box, where the local density remains relatively high. In these dense areas, ordered domains can span  the full height of the simulation box. In contrast, the low-density regions near the left and right boundaries of the box, consist primarily of disordered particles.  Between the dense crystal and the dilute phase, smaller crystalline domains are observed  drifting through the disordered background. This drift is attributed to  the slow, diffusion-driven expansion of the system, which gradually redistributes particles over time. Large ordered structures, similar to those shown in Fig. \ref{fig:plates_main}(b), persist throughout the entire duration of the simulations, which extended  up to $t=135 \tau_d$. These results  indicate  that the out-of-equilibrium  coexistence between ordered and disordered phases described in this section has a macroscopic lifetime of at least approximately $1.5 ~\mathrm{min}$, and potentially significantly longer. 

In addition, we also  consider conditions analogous to those in the  experimental study, where the system  equilibrated into a disordered fluid with a global packing fraction of $\eta_g = 0.02$. However, in our simulations at this global packing fraction, the crystal did not melt. This behavior is shown in the Supplementary Video  \cite{sm} and in Fig. \ref{fig:sup:v3_expansion.}(a), which shows the expansion of the crystal from simulation $P_7$. 
The color-coding of the particles denote the FCC class probability, determined using the method described in Section \ref{sec:fracFCC}, with principal components derived from simulation $P_6$.

From the final two frames of Fig. \ref{fig:sup:v3_expansion.}(a), it can be inferred that the particles have reached the boundaries of the simulation box. However, the system did not show any signs of melting throughout the entire simulation. This suggests that at a global packing fraction of $\eta_g = 0.02$, our simulations  do not reproduce the disordered equilibrium fluid observed experimentally by Larsen and Grier.\cite{larsen-1996,larsen-1997} This discrepancy suggests that the effective strength or range reported for the Yukawa  potential in Eq. (\ref{eq:hard_core_yukawa_potential}) in the experimental system may be overestimated.  

The first two frames of Fig. \ref{fig:sup:v3_expansion.}(a) show the system before the colloids reach the boundaries of the simulation box. During this early stage,  particles located near the dilute region are more likely to be assigned to the FCC class probability than particles  farther inward. This is an artifact of the classification method. Once the system has fully expanded and the particles form  a continuous structure--due to periodic boundary conditions--this artifact vanishes. Figure \ref{fig:sup:v3_expansion.}(b) displays the packing fraction gradient of simulation $P_7$ 
at $t=64\tau_d$. We observe that the only significant density variation occurs along the direction of expansion. 

\begin{figure}
    \centering
    \includegraphics[width=0.99\linewidth]{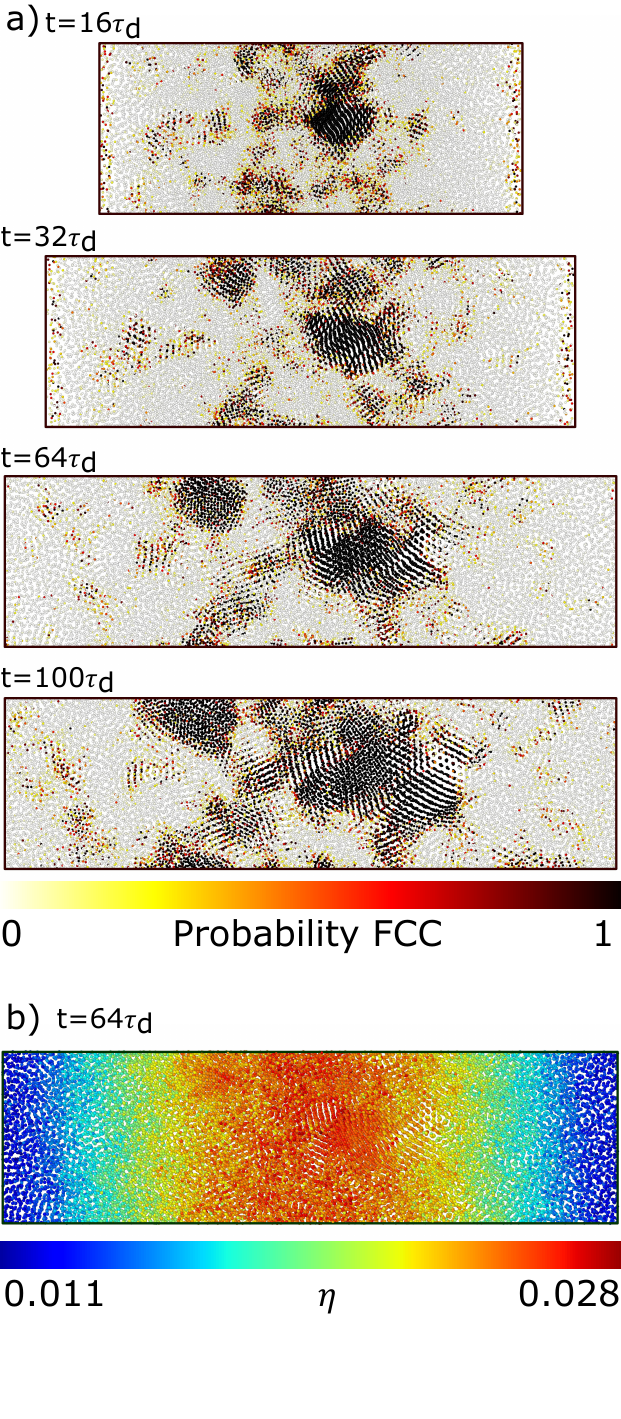}
    \caption{\textbf{Expansion of a crystal in a slab geometry at a global packing fraction $\eta_g=0.02$, corresponding to simulation $P_7$ of Table \ref{tab:plates}.} (a) The first four frames show the evolution of the crystalline regions during expansion of the crystal.  (b) Packing fraction gradient across the simulation box at time $t=64 \tau_d$, with the color bar indicating the local packing fraction $\eta$.}
    \label{fig:sup:v3_expansion.}
\end{figure}

While the shapes of the ordered regions in Fig. \ref{fig:plates_main}(b) and Figure \ref{fig:sup:v3_expansion.}(b) resemble those observed in the experimental  system, we did not observe any  significant density differences between the ordered and disordered regions. This suggest that additional factors are likely necessary  to explain the coexistence of  dense crystalline and dilute fluid regions reported by Larsen and Grier.\cite{larsen-1996, larsen-1997} 
Nonetheless, our simulations display long-lived order-disorder coexistence, characterized by islands of FCC-ordered particles that evolve in shape and drift as a result of the expansion of the system. These results indicate that not all features observed in  experiments necessarily stem  from like-charge attraction. Moreover, our findings highlight the strong effect of  system size and geometry on the stability and lifetime of superheated crystals. Although the compressed cubic crystallites  discussed in Section \ref{sec:results_cube} were initialized at nearly  ten times  the density of the crystal in the slab geometry, the latter maintained ordered regions for substantially  longer durations.

{\subsection{Walls and Like-Charge Attractions}}

One aspect not addressed in our analysis is the complex interplay between  walls and colloids. To place our long-box simulations in the broader context of wall-induced like-charge attraction, we  briefly discuss previous studies examining  electrostatic and hydrodynamic effects on the effective interactions of charged colloids near walls.

Many experimental studies have investigated the effective interaction potential between pairs of colloids, positioned either near a single plate or confined between two glass plates.\cite{polin-2007,baumgartl-2006,grier-2004,han-2003,franck-2003, brunner-2002,ramirez-saito-2003,behrens-2001, rao-1998, larsen-1997,crocker-1996, carbajal-tinoco-1996, kepler-1994} Several of these studies report  indications of like-charge attraction,\cite{polin-2007,grier-2004, han-2003, ramirez-saito-2003,rao-1998, crocker-1996, carbajal-tinoco-1996, kepler-1994, larsen-1997, brunner-2002} although possible experimental artifacts have also been raised.\cite{rao-1998, baumgartl-2006, polin-2007, gyger-2008} 
The most compelling evidence for an attractive potential between polystyrene particles arises in samples confined between two glass walls,\cite{grier-2004, han-2003, ramirez-saito-2003,rao-1998, crocker-1996, carbajal-tinoco-1996, kepler-1994} where strong confinement was found to be essential.\cite{grier-2004,han-2003, crocker-1996} Studies based on  Poisson-Boltzmann theory showed, however,  that the interaction potential between two like-charged colloids confined within a cylinder remains strictly repulsive.\cite{sader-1999, sader-1999-brief, neu-1999, trizac-1999} Importantly, both the strongly confined  geometry and the cylindrical confinement differ significantly from that considered in the present work.\cite{larsen-1996,larsen-1997} The effective interaction between colloids near a single wall has also been studied, where it was suggested that planes of colloids may generate partial confinement, allowing particles to act as walls for one another and transmit the influence of the container into the bulk crystal.\cite{larsen-1997,crocker-1996} While Larsen and Grier \cite{larsen-1997} reported an attractive interaction potential for polystyrene colloids near a single wall, these findings have been challenged by subsequent studies,\cite{grier-2004,franck-2003,brunner-2002} leaving the existence of such attractions for polystyrene colloids unclear. In contrast, attractive interactions have been more consistently observed for weakly acidic silica colloids.\cite{polin-2007}
Interestingly, for silica systems, the influence of a single wall decreases with decreasing ionic strength,\cite{polin-2007, behrens-2001, han-2003} whereas low-salt conditions are required for stabilizing metastable crystals.\cite{larsen-1996, larsen-1997} Thus, for polystyrene colloids, confinement-induced like-charge attractions have been observed only under strong confinement, whereas for silica colloids, the measured wall-induced attractions decrease with decreasing salt concentration—contrasting with the low-salt conditions required to maintain the metastability of superheated crystals. These observations suggest that confinement-induced attractions cannot explain the metastability  of superheated crystals formed by polystyrene latex colloids.

Finally,  Squires and Brenner \cite{Squires2000_brenner} showed that non-equilibrium hydrodynamic coupling between a pair of colloids located at a distance $h$ from a container wall can give rise to an apparent like-charge attraction. Their analysis assumed constant, uniform negative charges on all three objects, with the plate’s surface charge density treated as a fitting parameter. 

Notably, recent observations of like-charge attraction in samples of large membrane-coated colloids revealed an asymmetry with respect to the sign of the colloid charge: only negatively  charged colloids exhibited  particle clustering.\cite{baksh-2004, gomez-2009} Further experimental work showed that negatively charged particles can display like-charge attraction in polar solvents, whereas positively charged colloids do so in nonpolar solvents.\cite{wang-2024} Kubincová \textit{et al.} attributed this asymmetry to an electrosolvation effect,\cite{kubincova-2020} a conjecture that has since been supported by further studies reporting promising evidence.\cite{wang-2025, walker-gibbons-2022, behjatian-2022} Nonetheless, it remains an open question whether  the cases of like-charge attraction discussed here arise from  the same underlying mechanisms. \\

\section{Summary and conclusions}\label{sec:summary_and_conclusions}

In this study, we have performed simulations of charged  colloids, in which the colloid charge was considered as a dynamically evolving, density-dependent parameter. Using Poisson-Boltzmann cell theory calculations, we determined the renormalized and regulated charge as  functions of packing fraction. To capture the non-instantaneous charging behavior of colloidal surfaces, we introduced  a time-dependent differential equation governing the charging dynamics during the simulations. This approach was then applied to  low-salt suspensions of superheated colloidal crystals. To the best of our knowledge, this represents  the first simulation implementation of charge-regulating  colloids with explicitly density-dependent charges.  

The first series of simulations was initialised by placing a cubic colloidal crystal next to a colloidal fluid. Although all simulated crystals eventually melted, their lifetimes were significantly extended when both  density- and time-dependence were incorporated into the determination of  colloid charge and screening length. More specifically, the  simulations revealed two distinct regimes  based on the relaxation time  $\tau$, relative to the diffusion timescale $\tau_d$:  $\tau > \tau_d$ and $\tau < \tau_d$. Systems with $\tau > \tau_d$ exhibited notably longer  crystal lifetimes and are  most relevant for experimental settings where charge regulation is slow. In contrast, systems with $\tau < \tau_d$, where  the relaxation dynamics are effectively instantaneous, displayed faster melting dynamics. Within the context of our experimental model system, we observed  an out-of-equilibrium crystal-fluid coexistence lasting  at least $46~\text{s}$. Additionally, increasing the initial packing fraction of the crystal further enhanced its stability and lifetime.

In the second set of simulations, a colloidal crystal slab was positioned adjacent to a low-density colloidal gas phase, with the two phases separated by two planar interfaces as opposed to six interfaces in the case of a cubic crystal. As the crystal expanded, a dynamic coexistence emerged between  crystalline and fluid-like regions, with ordered crystalline domains drifting through surrounding disordered regions of colloids. These crystalline regions  grew in some areas and dissolved in others, leading to a continuously evolving structure. This non-equilibrium phase coexistence persisted throughout  the entire simulation duration of $t=135\tau_d$, closely resembling the heterogeneous structures observed by Larsen and Grier.\cite{larsen-1997} However, in contrast to the experimental observations, our simulations did not exhibit significant density differences between the crystalline  and fluid regions, as both phases maintained comparable local densities.  This discrepancy suggests that additional physical mechanisms are necessary to  fully account for the observed metastability and phase coexistence  in charged colloidal crystals. Furthermore, our results highlight the significant influence  of  system size and geometry on the stability and lifetime of superheated crystals. Although the compressed cubic crystallites  discussed in Section \ref{sec:results_cube} were initialized at nearly  ten times  the density of the crystal in the slab geometry, the latter sustained ordered regions for substantially  longer timescales.

Altogether, incorporating dynamically evolving, density-dependent colloidal interactions leads to unexpected dynamic melting behavior, characterized by long-lived coexistence between  crystalline domains and  disordered regions. These results underscore  the importance of conducting  more detailed simulations of systems with density-dependent charging, as such  interactions can give rise to non-intuitive and emergent out-of-equilibrium phenomena. Notably, our findings depend on the initial colloid configuration and the charge-regulation curve used. Changing the surface-charging mechanism under scrutiny is therefore expected to produce markedly different out-of-equilibrium responses, offering intriguing possibilities for tuning colloidal phase behaviour. 
{ Finally, we note that metastable crystalline clusters are often explained in terms of effective like-charge attractions. In contrast, our results show that the lifetime of metastable crystals can be significantly extended by the slow relaxation dynamics of surface charges—arising from charge regulation and charge renormalization—while the colloidal interactions remain purely repulsive.} 
{ It would be interesting to further investigate the kinetic relaxation mechanism presented here in combination with more detailed charge-regulating Poisson–Boltzmann calculations,  in order to capture the influence of  solvent effects and heterogeneous surface charge distributions on like-charged colloidal interactions.}

\section*{Author contributions}
\textbf{Laura Jansen:} Conceptualization (equal); Data curation (equal); Formal Analysis (lead); Investigation (lead); Methodology (equal); Software (lead); Visualization (lead); Writing - Original Draft (equal); Writing - Review \& Editing (supporting).
\textbf{Thijs ter Rele:} Conceptualization (equal); Data curation (equal); Investigation (supporting); Methodology (equal); Software (supporting); Visualization (supporting);  Writing - Original Draft (equal); Writing - Review \& Editing (supporting).
\textbf{Marjolein Dijkstra:} Conceptualization (equal); Funding Acquisition (lead); Methodology (equal); Project Administration (lead);  Supervision (lead); Writing - Review \& Editing (lead).

\section*{Conflicts of interest}
There are no conflicts to declare

\section*{Data availability}
The code used in generating and analyzing the data presented in work are included in the supplementary information. The data itself is not included, following from the large size of the generated files. However, it is readily available on request.


\section*{Acknowledgements}
M.D. and T.t.R acknowledge funding from the European Research Council (ERC) under the European Union’s Horizon 2020 research and innovation programme (Grant agreement No. ERC-2019-ADG 884902 SoftML). We thank Ren\'{e} van Roij for useful discussions.



\balance


\bibliography{main_abbreviated} 
\bibliographystyle{rsc} 

\end{document}


\appendix

{\centering

  {\large\bfseries Supplementary information: Surface Charge Relaxation Controls the Lifetime of Out-of-Equilibrium Colloidal Crystals\par}
  \vspace{1em}

  {\normalsize
    Laura Jansen\textsuperscript{1}, \;
    Thijs ter Rele\textsuperscript{1}, \;
    Marjolein Dijkstra\textsuperscript{1}
  \par}
  \vspace{0.5em}

  {\small\itshape
    \textsuperscript{1}Soft Condensed Matter \& Biophysics, Debye Institute for Nanomaterials Science, Utrecht University, Princetonplein 1, 3584 CC Utrecht, Netherlands
  \par}
  \vspace{2em}
}

\section{The Dissociation Boundary Condition}

The dissociation equilibrium at the colloid surface enters the Poisson-Boltzmann cell calculations via the charge-regulation boundary condition given by Eq. (6). In this section, we derive this relation to further elucidate the role of the bulk $\text{pH}$ and $\text{p}K_a$, where we follow the discussion as presented by Zoetekouw {\em et al.}\cite{zoetekouw-2006} We  define the variable $\alpha = Z/M$ to be the fraction of ionized surface sites. The dissociation constant $K_a$ for a certain reaction is typically known from tabulated data, making it a natural input parameter. This dissociation constant is defined as
\begin{equation}
    K_a = \frac{[\ce{A^-}]_s[\ce{H^+}]_s}{[\ce{AH}]_s}, 
\end{equation}
where $[\ce{A^-}]_s, [\ce{H^+}]_s$ and $[\ce{AH}]_s$ denote the \textit{surface} concentrations of  dissociated sites, hydrogen ions, and bound surface complexes, respectively, denoted by the subscript $s$.\cite{zoetekouw-2006} 
The local fraction of ionized sites can be defined as $\alpha =[\ce{A^-}]_s/\left( [\ce{AH}]_s +[\ce{A^-}]_s\right)$, which allows  the dissociation constant to be written  as
\begin{equation}
    \label{eq:k_a_simple}
    K_a = \frac{\alpha}{1-\alpha}[\ce{H^+}]_s.
\end{equation}
The surface concentration of hydrogen ions $[\ce{H^+}]_s$ is generally not directly accessible  experimentally, but within the Poisson-Boltzmann framework it is related to the Boltzmann distribution $[\ce{H^+}]_s = [\ce{H^+}] \exp{(-\phi(a))}$, where $\phi(a)$ is the dimensionless electrostatic potential at the colloid surface. Substituting this into  Eq. (\ref{eq:k_a_simple}) yields the charge-regulation relation 
\begin{equation}
    \alpha = \frac{1}{1+([\ce{H^+}]/K_a)\exp{(-\phi(a))}}.
\end{equation}
Using $\mathrm{p}K_{a} = -\log{(K_a)}$ and $\mathrm{pH} = -\text{log}([\text{H}^+])$, it is easily verified that this expression is equivalent to Eq. (6) in the main text.

\section{Poisson-Boltzmann Cell Theory Linearized at the Cell Boundary and the Constant Charge Case.}\label{sec:sup:edge_linearized_cell model_and_the_constant_charge}

\begin{figure*}[h]
\centering
\includegraphics[width=0.99\textwidth]{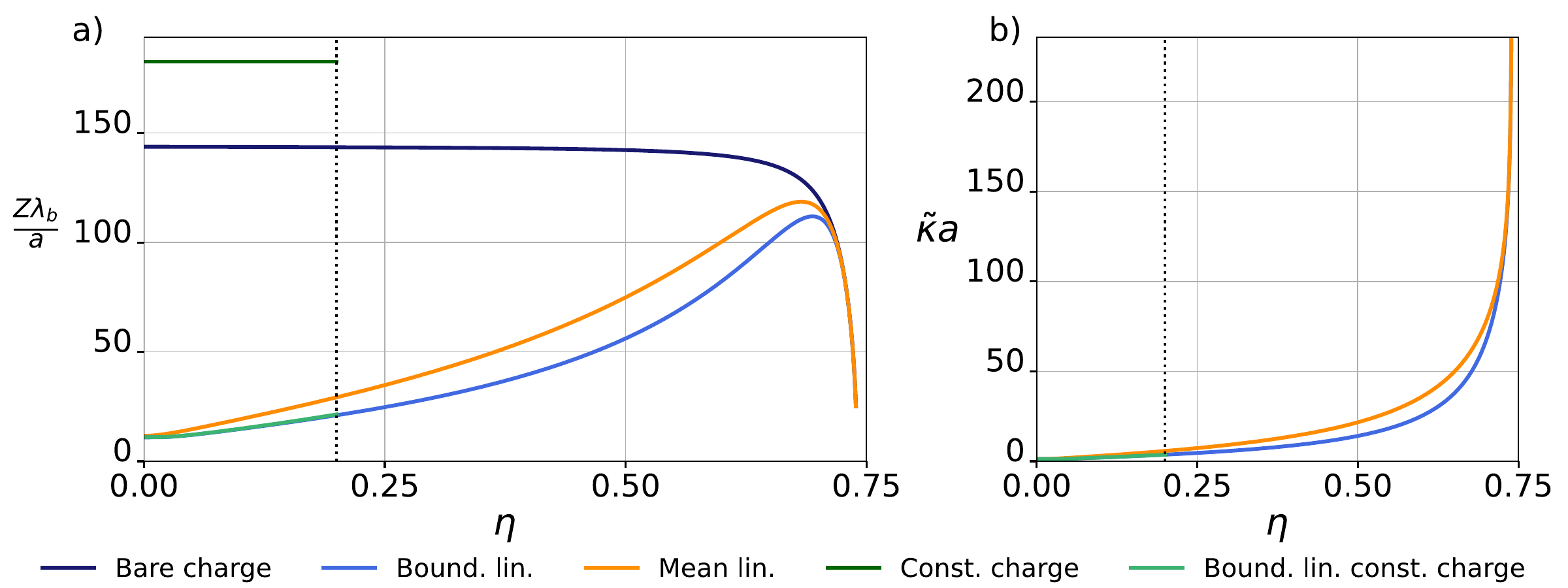}
\caption{\label{fig:supplementary_info_renormalization}\textbf{Charge and Inverse Screening Length.} (a) Bare charge (dark blue), renormalized charge by linearizing at the cell boundary (light blue) or around the mean electric potential (orange)   for colloids with  $M = 10^5$ surface groups, radius $a = 465.5 \lambda_B$, inverse screening length $\kappa = 1.185/a$ as taken from Ref. \citenum{larsen-1996, larsen-1997}, and  $[\text{H}^+]/K_{a} = 0.05$, along with the colloid charge (dark green) for colloids with a constant charge boundary condition ($Z = 8.5 \cdot 10^{4}$) and its renormalized charge (light green) by linearizing at the cell boundary, which were numerically stable only until $\eta \approx 0.2$, see the dotted line, and (b) corresponding inverse screening lengths using the same color scheme.}
\end{figure*}

In the main text, we mentioned that the Poisson-Boltzmann (PB) equation can be linearized about  different reference points \cite{brito-2023}. In this work, we chose to linearize the PB equation around the  mean electric potential $\phi_m$.  One advantage of this approach is that it extends the range  of  low packing fractions $\eta$ over which a one-to-one mapping exists  between the renormalized charge $Z^*$ and the renormalized inverse screening length $\tilde{\kappa}$. The more common convention, however,  is to linearize the PB equation at the cell boundary $R$. The  linearization at the cell boundary $R$ leads to  the following expressions for the renormalized charge   
\begin{eqnarray}
\label{eq:Trizac_effective_charge_cell_edge}
 Z^*=&&-\frac{\tanh{\phi(R)}}{\tilde{\kappa}\lambda_B} \left\{  (\tilde{\kappa}^2aR -1)\sinh{[\tilde{\kappa}(R-a)]}\right. \nonumber\\ 
 && 
 + \left.\tilde{\kappa}(R-a)\cosh{[\tilde{\kappa}(R-a)]} \right\},
\end{eqnarray}
and renormalized inverse screening length  \cite{trizac-2003}
\begin{equation}
    \tilde{\kappa}^2=\kappa^2\cosh{\phi(R)}.
\end{equation}
{ As mentioned in Section 4.1, our choice of the ratio  $[\text{H}^+]/K_{a}$ was constrained by  the numerical stability of our calculations. 
Assuming the proton concentration equals  the reported salt concentration $[\text{H}^+] = 5 ~\si{\micro \molar}$ \cite{larsen-1997},  our chosen ratio $[\text{H}^+]/K_{a} = 0.05$  corresponds to $\mathrm{p}K_{a} = 4.0$. Small amounts of stray salt would alter this value only slightly.} 
Nonetheless, this value underestimates the acidity of our model system, which is based on polystyrene latex particles with strongly  acidic sulphate groups. A lower (more realistic) $\mathrm{p}K_{a}$ would  yield a higher bare colloid charge. To demonstrate  that the PB equation is not strongly affected by the magnitude of the bare colloid charge in this  parameter regime, we also calculated the renormalized charge as a function of packing fraction $\eta$ for colloids with a fixed bare  charge of $Z = 8.5 \cdot 10^{4}$. This calculation remained numerically stable up to a packing fraction of approximately  $\eta\approx 0.2$. 

In Fig.~\ref{fig:supplementary_info_renormalization}(a), we present the bare colloidal charge for the system studied in the main text, which was based on a simple dissociation boundary condition with $M=10^5$ surface groups, a colloid radius $a=465.5 \lambda_B$, inverse screening length $\kappa  =1.185/a$ and $[\text{H}^+]/K_{a}=0.05$. The bare charge $Z$, plotted  as a function of packing fraction, is shown by the dark blue line in Fig. \ref{fig:supplementary_info_renormalization}(a) and corresponds to  the same curve discussed  in the main text. In addition, we also plot the renormalized charge $Z^*$ as obtained by linearizing around the mean electric potential $\phi_m$ (orange),  as used in the main text, and  by linearizing at the cell boundary $R$ (light blue). We also include results for a fixed bare charge of $Z = 8.5 \cdot 10^{4}$ (dark green), along with its corresponding renormalized charge using  linearization at the cell boundary $R$ (light green). In  Fig. \ref{fig:supplementary_info_renormalization}(b), we show the corresponding renormalized inverse screening lengths $\tilde{\kappa} $ using the same color scheme.

Given that the light blue and light green curves in Fig. \ref{fig:supplementary_info_renormalization}(a) nearly overlap, we conclude that our choice of $[\text{H}^+]/K_{a}$ has little effect on  our final results. Interestingly, the point of linearization has a  significant impact on  the magnitude of the renormalized charge, as can be inferred from the large difference between the  orange (mean electric potential) and light green (cell boundary) curves in Fig. \ref{fig:supplementary_info_renormalization}(a). This is expected, since  linearization around the mean electric potential inherently yields  a higher renormalized charge and inverse screening length than linearization at the lower electric potential at the cell boundary \cite{brito-2023}.

{
\section{Concentration Dependence of the Surface Charge Relaxation Time}\label{sec:tau_concentration}
In the main text, we made the assumption that the rate constant $k$ governing the charging dynamics in Eq. (11) is constant, and consequently that the relaxation timescale  $\tau$ in Eq. (12) is also constant.  However,  the relaxation time $\tau$ is in fact expected to depend on the counterion concentration $\left[\mathrm{H^+}\right]$, since counterion concentration is a major determinant in charge regulation processes.  
A more physical description of the regulated colloid charge dynamics that depends on counterion concentration  can be written as
\begin{equation}
        \frac{\partial Z(t)}{\partial t} = M L_\zeta \left( \ln{K_a}  -\ln{\left(\frac{(Z/M)}{1-(Z/M)}\left[\mathrm{H^+}\right] \right)}\right),
\end{equation}
where $L_\zeta$ is the  kinetic coefficient.\cite{takae-2018} Expanding this equation to first order around the equilibrium charge $Z(t) = Z_{eq}$ yields
\begin{equation}
    \frac{\partial Z(t)}{\partial t} =  \frac{ L_\zeta }{(1 - Z_{eq}/M)(Z_{eq}/M)} \left( Z_{eq} - Z(t)\right),
    \label{eq:zeta_expanded}
\end{equation}
where we used  the mass action law to define the dissociation constant $K_a = \left[\mathrm{H^+}\right] (Z_{eq}/M)/(1-Z_{eq}/M) $. By comparing Eq.~(12) and Eq. (\ref{eq:zeta_expanded}), and using $\left(Z_{eq}/M\right)/\left( 1- Z_{eq}/M\right) = K_a/\left[\mathrm{H^+}\right]$, we obtain 
\begin{equation}
       \tau =\frac{1}{L_\zeta} \frac{\left[\mathrm{H^+}\right]/K_{a}}{\left(1 + \left[\mathrm{H^+}\right]/K_{a}\right)^2}.
\end{equation}
This expression shows that, within the low counterion concentration regime studied here, $\tau$ increases with $\left[\mathrm{H^+}\right]$. Nevertheless, the order of magnitude of $\tau$ is primarily determined by $L_\zeta$, which  depends strongly on the specific colloidal system under consideration, as discussed in Section 2.3. 
Although it would be interesting to  account for the counterion-concentration dependence of $\tau$, its dominant contribution---and thus the governing factor to the colloid dynamics---remains determined by $L_\zeta$.
This kinetic coefficient $L_\zeta$ still largely determines the relevant characteristic timescale $\tau$ and  thereby the resulting colloid dynamics.}

{ \section{Time Evolution of the Effective Charge and the Dimensionless Screening Parameter}\label{sec:time_evolution_supp}

One of the main findings of this work is that the renormalized charge $Z^*$ and reduced screening parameter  $\tilde{\kappa}a$ are dynamical quantities that evolve over the lifetime of an out-of-equilibrium system. We  propose a relaxation mechanism for this process, discussed in detail  in the main text. Specifically, the renormalized charge $Z^*$ is updated according to Eq. 12, while the equilibration of $\tilde{\kappa}a$ follows from the numerical relation between these two quantities. 

Here we analyze simulations $B_{9}$ to $B_{11}$, which differ only in their values of $\tau$, see Table 1. Two representative particles were selected to illustrate the characteristic behaviour exhibited by colloids in different regions of the system. The red particle in Fig. \ref{fig:supplementary_info_time_evolution} is situated at the interface between the superheated crystal and the dilute fluid, while the orange particle is initially located in the interior of the crystal. The time evolution of $Z^*\lambda_b/a$ and $\tilde{\kappa}a$ for the red and orange particles are depicted in Figs.~\ref{fig:supplementary_info_time_evolution}(b) and (c), respectively. 

The interface particle hardly experiences a change in its local packing fraction; consequently, its equilibration is dominated by Brownian-motion-induced fluctuations in the packing fraction, reflected in the  erratic  curves of Fig.~\ref{fig:supplementary_info_time_evolution}(b). Moreover, the curves are largely insensitive to the value of $\tau$. In contrast, the interior particle undergoes a smooth, strongly $\tau$-dependent relaxation as shown in  Fig.~\ref{fig:supplementary_info_time_evolution}(c). This behavior stems from the  large initial drop in  local packing fraction. These results  highlight  the strong dependence of the system's evolution  on its initial configuration.

\begin{figure*}
\centering
\includegraphics[width=0.99\textwidth]{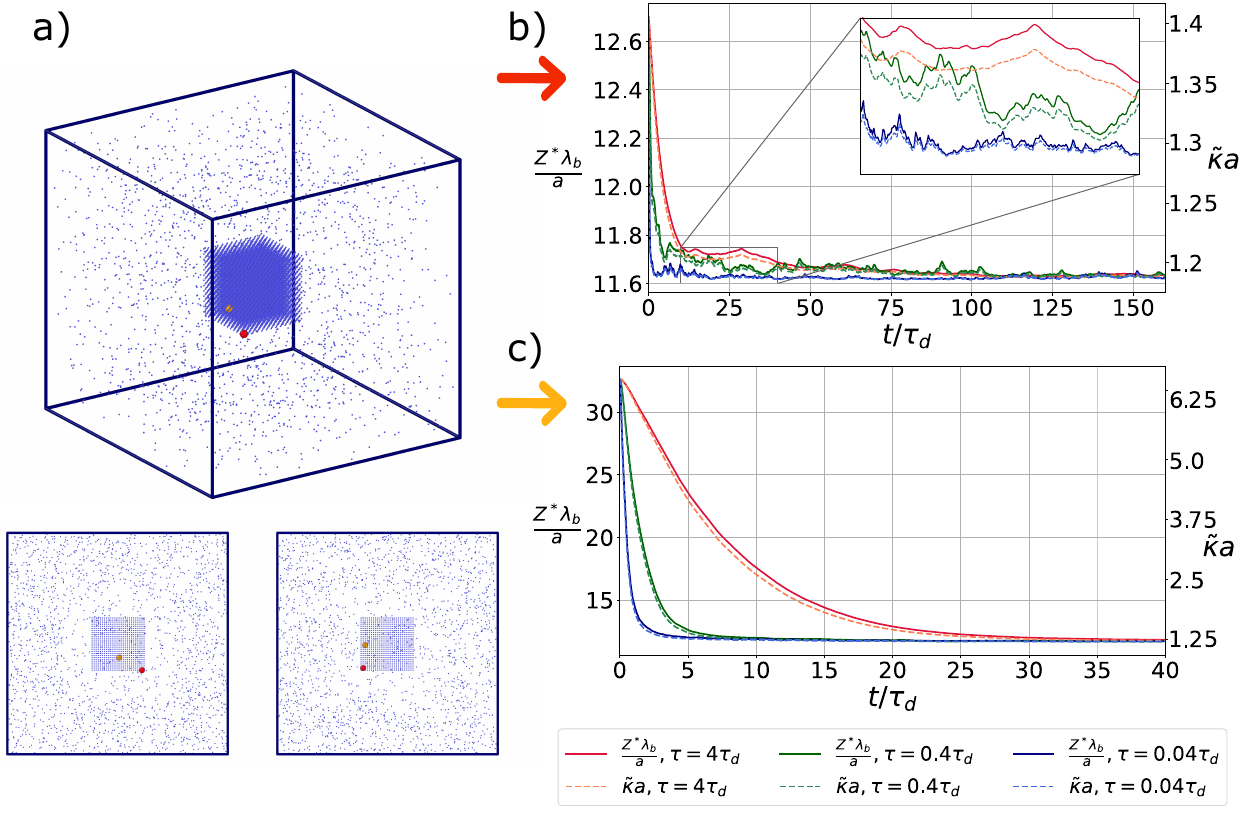}
\caption{\label{fig:supplementary_info_time_evolution}\textbf{Time evolution of the effective charge and inverse screening length for particles in the interior (orange) and at the exterior (red) of the superheated crystal.} (a) Initial configuration of the superheated crystal, corresponding to simulations  $B_{9}$-$B_{11}$. Particles are depicted with a radius of $0.3\sigma$ for visualization purposes. Two projections of the system are included, one through the front-left plane (left) and one through the upper plane (right). Evolution of the effective charge and screening length for two representative  particles (radius $1.5\sigma$): one located at the outer shell (red) of the superheated crystal in panel (b) and one in the interior (orange) in panel (c).  The inset of panel (b) shows the effect of local density fluctuations on the effective parameters. This effect is absent in panel (c), where the large initial drop in local packing fraction overshadows the smaller density fluctuations caused by Brownian motion.}
\end{figure*}

As mentioned in the main text, the effect of the relaxation mechanism can be viewed as allowing for alternate trajectories in the space spanned by $Z^*\lambda_b/a$, $\tilde{\kappa}a$ and $\eta$. In the instantaneous limit, these trajectories coincide with the orange lines in Fig.~\ref{fig:supplementary_info_renormalization}. Figs.~\ref{fig:supplementary_info_eta_v_charge_evolution_eta_space}(a) and (b) show the trajectories for the red and orange particles in  Fig.~\ref{fig:supplementary_info_time_evolution}(a). For clarity, only the effective charge is shown; however,  the evolution of the reduced screening parameter $\tilde{\kappa}a$ closely mirrors that of $Z^*\lambda_b/a$. Clearly, a large relaxation timescale $\tau$ effectively ``freezes'' the evolution of  these dynamic quantities. Furthermore,  $Z^*\lambda_b/a$ approaches the instantaneous relaxation limit from above, because of the compressed nature of the superheated crystal. 

\begin{figure*}
\centering
\includegraphics[width=0.99\textwidth]{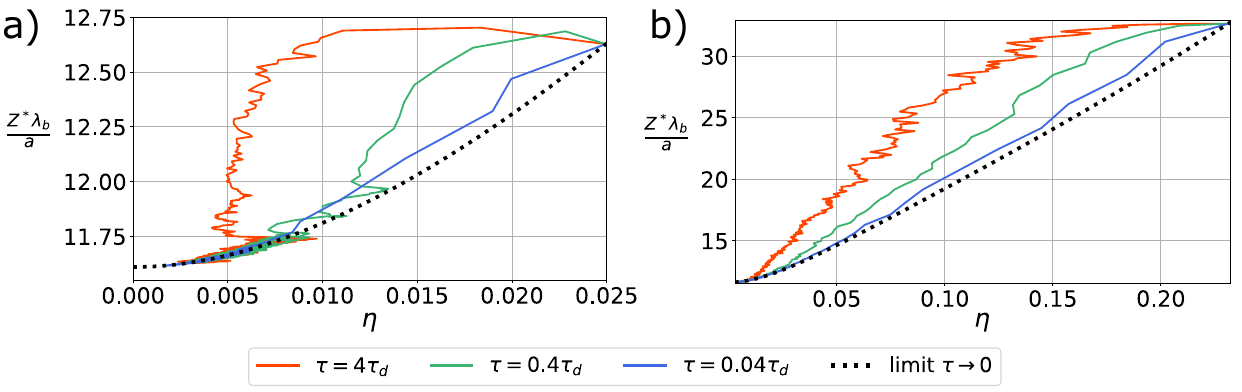}
\caption{\label{fig:supplementary_info_eta_v_charge_evolution_eta_space}\textbf{Influence of the relaxation time on the trajectory of the effective charge as a function of packing fraction.} Panels (a) and (b) correspond to the red and orange particles depicted in Fig. \ref{fig:supplementary_info_time_evolution}(a), respectively. The introduction of a relaxation time via Eq. 12 causes the effective charge to approach the instantaneous curve (dotted line), which corresponds to the orange line in Fig. \ref{fig:supplementary_info_renormalization}(a). The resolution of the data in this figure is limited by the frequency at which values were recorded in the simulations.}
\end{figure*}
}

\section{Identification of Crystalline Particles}\label{sec:crysfrac}

\subsection{Principal Component Analysis and Gaussian Mixture Model}
For each particle and timestep, we calculate  10 BOPs (Eq. (17) with $l\in[2, 11]$). 
Our input data consists of a $10 \times N$ matrix for each  simulation timestep, where $N$ is the number of particles. We concatenate  these matrices across all timesteps and applied Principal Component Analysis (PCA) to reduce the dimensionality of the data.  PCA is a commonly used technique  for dimensionality reduction and feature extraction, which identifies new axes (principal components) that capture the directions of maximum variance in the data  \cite{bishop2006pattern}. The goal is to project the data onto a  lower-dimensional space while preserving its most  important features. We used the PCA implementation provided by  \textsc{Sklearn}  \cite{JMLR:v12:pedregosa11a}. The principal components were calculated for a predetermined training set and were subsequently applied to the test set. We selected the first  two principal components to enable straightforward visualisation of the data.

Next, we applied  a Gaussian Mixture Model (GMM) to the projected data obtained from PCA. Mixture models, such as the GMM,  construct complex probability distributions by combining simpler ones---in this case, multivariate Gaussian distributions \cite{bishop2006pattern}. GMMs  are commonly used for clustering  data, which is also our aim here \cite{bishop2006pattern}. By using a GMM, we implicitly assume that the data can be represented  as a weighted combination of Gaussian distributions. To fit the model, we used the expectation-maximization algorithm  to find the maximum likelihood estimates of the Gaussian parameters,   \cite{bishop2006pattern} as implemented in \textsc{Sklearn} \cite{JMLR:v12:pedregosa11a}. To determine the optimal number of Gaussian components, we employed the Bayesian Information Criterion (BIC) \cite{schwarz-1978, bishop2006pattern},  again using \textsc{Sklearn} \cite{JMLR:v12:pedregosa11a}. In all cases, the BIC curve exhibited  a clear elbow at two components, indicating that a  two-component model is a sensible choice. This allowed us to interpret one Gaussian component as corresponding to particles in the FCC crystal phase and the other to those in the fluid phase. For each particle, we computed the class probability of belonging to  the crystal phase. These probabilities were grouped back into individual timesteps, and the sum of crystal probabilities across all particles in a single timestep was taken as a measure of the  fraction of crystallized particles. Throughout this  workflow, the data was normalized to have zero mean and unit variance  before applying PCA and GMM.

\begin{figure*}
    \centering
    \includegraphics[width=0.99\linewidth]{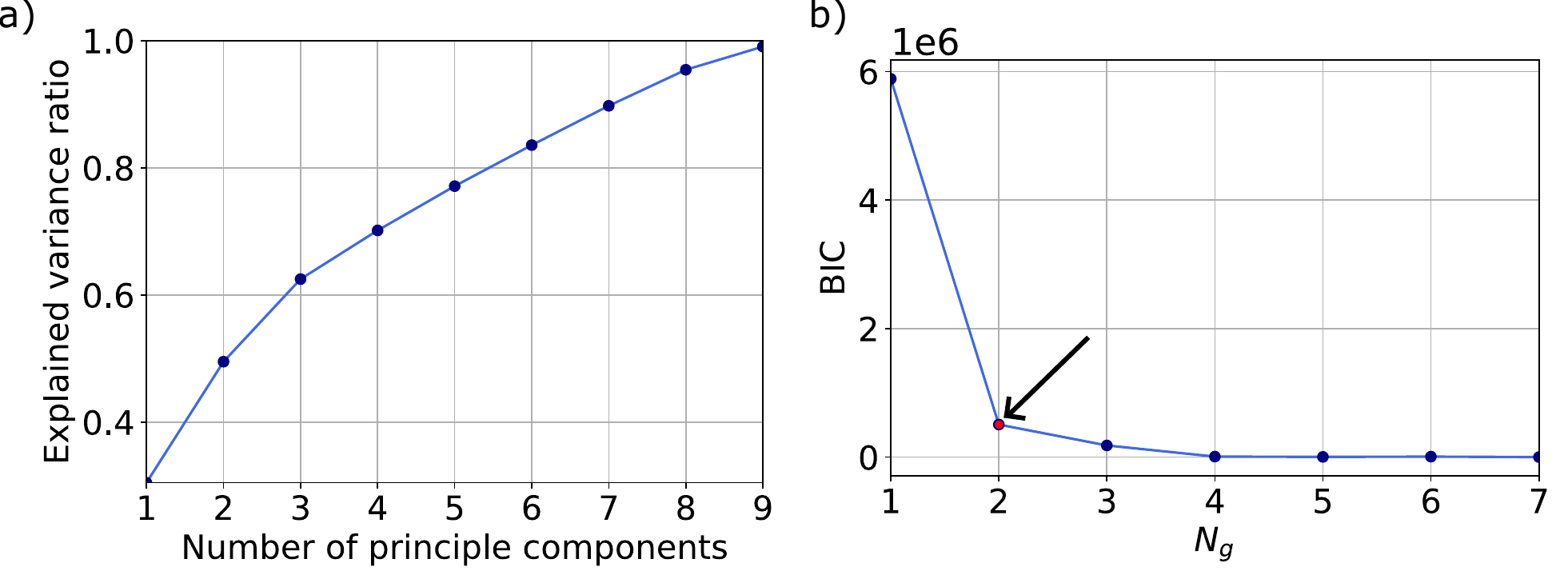}
    \caption{\textbf{Principal Component Analysis (PCA) and Gaussian Mixture Model (GMM) for a Cubic Crystallite.} (a) Explained variance as a function of the number of principal components for PCA using simulation $B_{12}$ as training data (b) and the Bayesian Information Criterion (BIC) as a function of the number of Gaussian components $N_g$ for the GMM trained on data from $B_{9}$. This data was projected onto the two principal components  obtained from $B_{12}$. The BIC as a function of the number of Gaussian components  shows an elbow at $N_g=2$ (red dot), indicating this is a viable choice.}
    \label{fig:sup:BIC_and_PCA}
\end{figure*}

We apply PCA on the dataset obtained from simulation $B_{12}$ (see Table 1). 
Fig. \ref{fig:sup:BIC_and_PCA}(a) shows the explained variance ratio as a function of the number of principal components. Projecting the $10$ BOPs per particle onto just two principal components retains approximately  $50\%$ of the total variance in  the data, while also allowing for straightforward visualisation. 

These principal components were subsequently used to reduce the dimensionality of the data for configuration $B_{9}$. A Gaussian Mixture Model was trained on this reduced dataset while varying the number of Gaussian components $N_g$.  As shown in Fig. \ref{fig:sup:BIC_and_PCA}(b), the BIC value exhibits a clear elbow at $N_g=2$, supporting the choice of   two  components. However, we note that the BIC curve does not display a well-defined minimum, indicating that the data may not be optimally described by a simple mixture of Gaussian distributions. More sophisticated approaches, such as those combining multiple Gaussian components to form non-Gaussian clusters (e.g. Baudry \textit{et al.} \cite{baudry-2010}), were also tested. However, these methods did not yield  significant improvements in classification accuracy while introducing additional complexity, and were therefore not pursued further.

\subsection{Different PCA and GMM training sets}

In the main text, we extensively discussed Fig. 2, which presents data from configurations $B_{9}$,  $B_{10}$, and $B_{11}$. The principal components used for dimensionality reduction were derived from configuration $B_{12}$, which differs from these three configurations only in the relaxation time $\tau$. In this section, we show that the choice of training data, for both PCA and GMM, has minimal effect on the final results.

\begin{figure}[hb]
    \centering
    \includegraphics[width=0.69\textwidth]{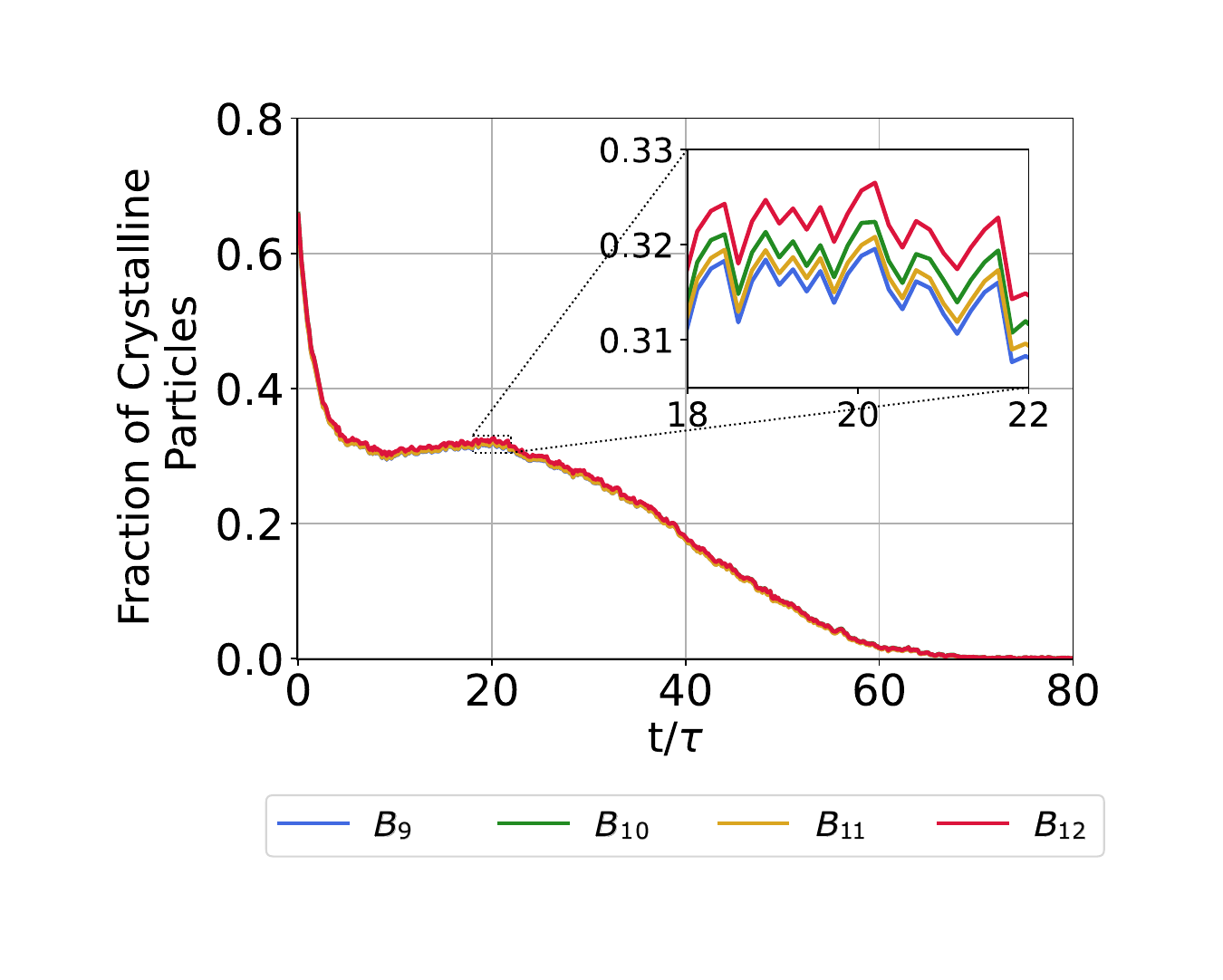}
    \caption{\textbf{Effect of different PCA training sets.} Fraction of crystalline particles as a function of time $t/\tau_d$ for simulation $B_{9}$ using PCA and GMM trained on four different training sets obtained from simulations $B_{i}$ for $i\in\{9,10, 11, 12 \}$.   The legend indicates which dataset was used for the PCA training, while the inset shows the small differences between the different curves.}
    \label{fig:sup:different_pca_training_effect_combi_fig}
\end{figure}

We  first examine how the choice of  PCA training data affects the fraction of crystalline particles. To this end, we projected the BOPs from simulation $B_{9}$ onto four different sets of principal components, each obtained by training PCA on a different configuration: $B_{9}$ (blue), $B_{10}$ (green), $B_{11}$ (yellow) and $B_{12}$ (red). These configurations differ only in their relaxation time $\tau$. After projection, we trained a separate GMM on each of the four reduced datasets and used the resulting models to calculate the fraction of crystalline  particles in $B_{9}$, as described earlier.

We clearly observe that the resulting curves for the fraction of crystalline particles are nearly indistinguishable and overlap closely, as highlighted in the inset of Fig.~\ref{fig:sup:different_pca_training_effect_combi_fig}. We therefore conclude that the data analysis is robust with respect to variations in the PCA training data. In this case, the impact on the fraction of crystalline particles is minimal—on the order of $1\%$. 

\begin{figure}[hb]
\includegraphics[width=0.69\textwidth]{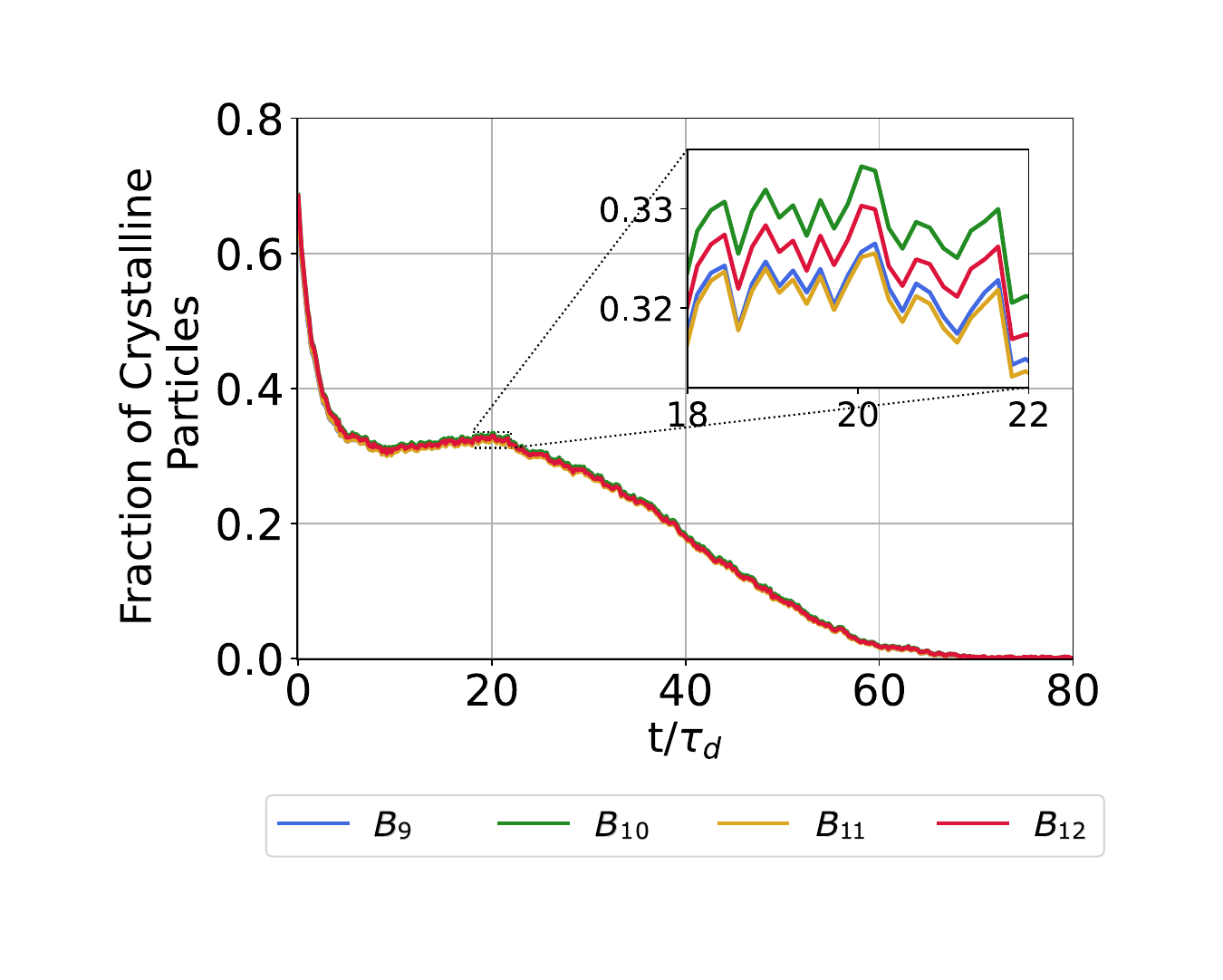}
\caption{ \textbf{Effect of different GMM training sets.} Fraction of crystalline particles as a function of time $t/\tau_d$ for simulation $B_{9}$ using PCA trained on data from simulation $B_{9}$ and GMM trained on four different training sets obtained from simulations $B_{i}$ for $i\in\{9, 10, 11, 12\}$.   The legend indicates which dataset was used for the GMM training, while the inset shows the small differences between the different curves.
} 
\label{fig:sup:difference_training_GMM_combi_fig}
\end{figure}

In addition, we trained four different GMMs using datasets obtained from simulations $B_{i}$ for $i\in\{9,10, 11, 12\}$ projected on the principal components obtained from $B_{12}$. These four different GMMs were subsequently used to distinguish two clusters in our test set $B_{9}$. The resulting fraction of crystalline  particles is shown in Fig.~\ref{fig:sup:difference_training_GMM_combi_fig} for all four GMMs: $B_{9}$ (blue), $B_{10}$ (green), $B_{11}$ (yellow) and $B_{12}$ (red). The difference between the fraction of crystalline particles obtained using these four different GMMs is on the order of $1\%$. Thus, we establish that our conclusions concerning the time it takes for the cubic crystal to melt is robust with respect to  variations in the training data.

\subsection{Interpretation of the Gaussian clusters}

In this section, we analyze the structure of the reduced BOPs to assess how well the Gaussian components correspond to  fluid and FCC phases. We focus again on simulation $B_{9}$, projecting its data onto the principal components trained on $B_{12}$, and apply the GMM to the test data from $B_{9}$. 

\begin{figure*}
\centering
\includegraphics[width=0.98\textwidth]{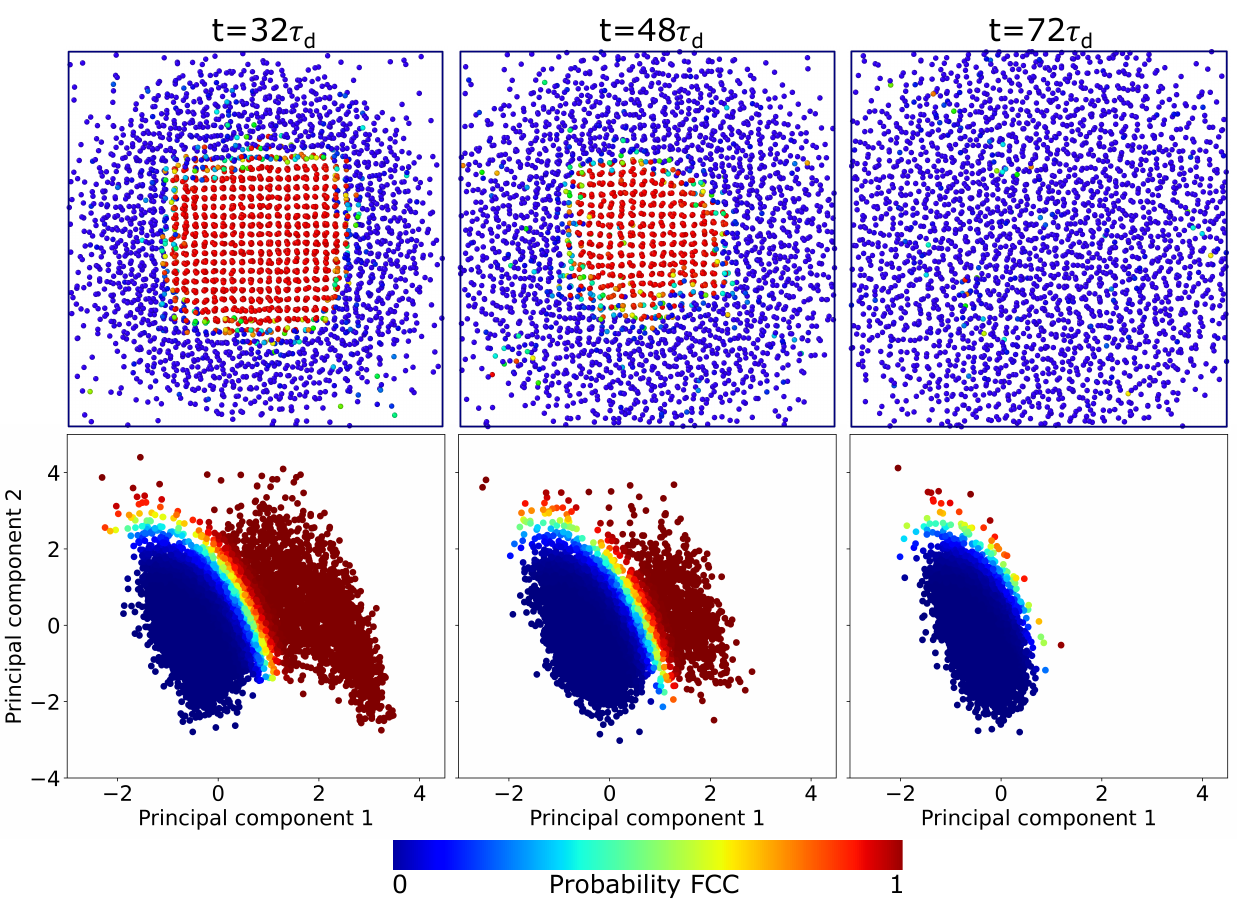}
\caption{\textbf{Identification of Crystalline Particles during  Melting of a Cubic Crystallite.} (Top) Typical configurations of a slice with thickness $20\sigma$ centered around the origin from simulation $B_{9}$ at three different time points. Particles are colored by their class probability of belonging to the FCC cluster. Red indicates a probability of one (fully FCC-like), while blue indicates zero probability (fully fluid-like), see colorbar. For visual clarity, the particle radius is set to $0.8\sigma$. (Bottom) Projection of the BOPs onto the first two principal components at the same time points.  The blue cluster corresponds to fluid-like particles, while the red brown cluster represents  FCC-like particles, where the latter progressively shrinks over time, indicating melting of the crystal. 
}\label{fig:sup:PCA}
\end{figure*}

In the top row of Fig. \ref{fig:sup:PCA}, we show typical configurations from  simulation $B_{9}$, with particles colored by their class probability for belonging to the FCC cluster. As time advances, the number of particles identified as FCC progressively decreases. The bottom row depicts the corresponding reduced BOPs, projected onto two principal components. From these projections, it becomes apparent that the blue oval represents the fluid particles; this cluster remains present even after the system has fully melted. In contrast, the brown red structure, representing the FCC phase, gradually merges with the fluid cluster and loses its distinct shape, indicating a continuous transition of FCC particles into a more fluid-like state.   This figure also underpins the conclusion that it is reasonable to define the fraction of crystalline particles as zero at late times. Although  the snapshot at $t=72 \tau_d$ still shows some orange and red particles, visual inspection confirms that they no longer form  any meaningful FCC structure. We therefore conclude that this straightforward classification  method is effective in capturing the global phase behavior of the system.

\newpage
\bibliography{supplementary} 
\bibliographystyle{rsc}